\documentclass[11pt,a4paper]{article}
\usepackage[margin=1in]{geometry}

\usepackage[utf8]{inputenc}
\usepackage{graphicx}
\usepackage{xcolor}
\usepackage{microtype}

\usepackage{bbm}
\usepackage{amsmath,amssymb,amsfonts,amsthm}
\usepackage{siunitx}
\usepackage{booktabs}

\usepackage{hyperref}

\theoremstyle{remark} 
\newtheorem*{remark}{Remark}

\renewcommand{\vec}[1]{\boldsymbol{#1}}

\newcommand{\dd}{\mathrm{d}}
\newcommand{\pfrac}[2]{\frac{\partial #1}{\partial #2}}

\newcommand{\veps}{\vec{\varepsilon}}
\newcommand{\vepsref}{\veps_\text{ref}}
\newcommand{\heaviside}[1]{H\left(#1\right)}
\newcommand{\ctensor}{\mathbbm{C}}

\author{Sebastian D.~Proell, Wolfgang A.~Wall, Christoph Meier}
\title{A simple yet consistent constitutive law and mortar-based layer coupling schemes for thermomechanical macroscale simulations of metal additive manufacturing processes}

\begin{document}
\maketitle
\begin{abstract}
    This article proposes a coupled thermomechanical finite element model tailored to the macroscale
    simulation of metal additive manufacturing processes such as selective laser melting. A first focus lies on the derivation of a consistent constitutive law on basis of a Voigt-type spatial homogenization procedure across the relevant phases, powder, melt and solid. The proposed constitutive law accounts for the irreversibility of phase change and consistently represents thermally induced residual stresses. In particular, the incorporation of a reference strain term, formulated in rate form, allows to consistently enforce a stress-free configuration for newly solidifying material at melt temperature. Application to elementary test cases demonstrates the validity of the proposed constitutive law and allows for a comparison with analytical and reference solutions. Moreover, these elementary solidification scenarios give detailed insights and foster understanding of basic mechanisms of residual stress generation in melting and solidification problems with localized, moving heat sources. As a second methodological aspect, dual mortar mesh tying strategies are proposed for the coupling of successively applied powder layers. This approach allows for very flexible mesh generation for complex geometries. As compared to collocation-type coupling schemes, e.g., based on hanging nodes, these mortar methods enforce the coupling conditions between non-matching meshes in an $L^2$-optimal manner. The combination of the proposed constitutive law and mortar mesh tying approach is validated on realistic three-dimensional examples, representing a first step towards part-scale predictions.
\end{abstract}

\section{Introduction}
Metal additive manufacturing (AM) opens new opportunities in manufacturing technology~\cite{Gibson2015,Herzog2016}. Specifically, the family of powder bed fusion additive manufacturing (PBFAM) processes allows for high geometrical flexibility and the potential for a pointwise control of material properties. However, many of the underlying complex multiphysics phenomena are still insufficiently understood and a suboptimal choice of processing parameters might lead to poor part quality in terms of pores or high residual stresses, which in turn might induce cracks in the part~\cite{Meier2017}. Currently, extensive process tuning (via trial and error) is required, which severely hinders adoption by industry. This limitation could be overcome via adequate models that allow for sound and real physics-based simulation of the whole process in order to predict optimal processing parameters. Large efforts have been undertaken to establish such approaches, all of which still face a number of obstacles when it comes to manufacturing of realistic and complex parts.

Existing modeling approaches for PBFAM are often distinguished by the modeled length scales (see overview in \cite{ Meier2021physics, Meier2017}). Macroscale or part-scale models determine residual stresses or dimensional warping. Residual stresses are of special interest in the community and a general review and theoretical considerations can be found in~\cite{Li2018a, Mercelis2006}. Mesoscale models operate on a length scale from single powder particles up to one powder layer thickness in order to study mechanisms of defect generation arising from the melt pool \cite{Khairallah2014,Korner2011,Markl2015,Meier2021novel,Russell2018,Wessels2018,Yan2018} or the powder recoating process \cite{Herbold2015,Meier2019a,Meier2019}. Microscale models investigate the formation of (typically anisotropic) metallurgical microstructures during solidification \cite{Gong2015,Lindgren2016,Nitzler2021,Rai2016,Salsi2018,Zhang2013}.

In this work we propose a coupled thermomechanical macroscale model based on the finite element method (FEM). The focus lies on two important aspects, namely consistent constitutive modeling in a homogenized macroscale sense and elaborate layer coupling schemes allowing for layerwise non-matching finite element (FE) meshes.
Macroscale models commonly solve a thermomechanical (or sometimes a pure thermal) problem, where powder, melt and solid phase are all modeled as homogenized continua with specific temperature- and phase-dependent thermal and mechanical  properties~\cite{Cervera1999,Childs2005,Denlinger2015,Hodge2014,Hodge2016,Kollmannsberger2019, Roy2018,Shen2012}. The heat of the incident laser beam is frequently modeled with the powder bed absorption model from Gusarov et al.~\cite{Gusarov2008,Gusarov2005,Gusarov2009}.
Further approaches that are frequently followed for an efficient simulation of PBFAM processes include spatial mesh adaptivity~\cite{Kollmannsberger2017,Riedlbauer2017,Denlinger2014,Zhang2018a}, sometimes combined with code parallelization and load balancing techniques~\cite{Neiva2019}, reduction of the computational domain via equivalent thermal boundary conditions representing the powder phase~\cite{Neiva2020}, process layer agglomeration~\cite{Zaeh2010,Hodge2016,Zhang2018a}, as well as rather heuristic approaches such as inherent strain schemes~\cite{Keller2014, Li2018, Setien2019}.

In macroscale models, all three phases are typically modeled in a Lagrangian manner based on a quasi-solid constitutive law with artificial stiffness values in powder and melt phase that are orders of magnitude lower than the stiffness of the solid phase. This approximation is well-justified since the mechanical stresses arising in the powder and melt phase at the stress-free surface of each processed layer are typically very small compared to the stresses in the bulk solid material. The interfaces between these different phases are commonly modeled as diffuse interfaces, e.g., defined by the solidus and liquidus temperature of the alloy.

Critically, such modeling approaches have to ensure that comparatively large strains, as typically arising in the melt phase (with low stiffness), do not transfer to large stresses in the newly formed solid phase (with high stiffness), which is assumed to be (approximately) stress-free right after solidification. Some contributions~\cite{Hodge2014, Li2018} use a plastic constitutive law, which can achieve the desired effect of an (almost) stress-free configuration at solidification start~\cite{Kruth2012}. Other works mention a reset or annealing of (plastic) strains and stresses if material (re)melts \cite{Bruna-Rosso2020,Denlinger2015, Parry2016}. Unfortunately, most existing works do not go into detail how a material model, that was initially designed for a single solid phase, is applied to a three-phase mixture of powder, melt and solid, and how the aforementioned condition of a stress-free solidification start can be incorporated in such models.
A notable exception are the recent contributions~\cite{Bartel2018,Noll2020}, where a constitutive model for the three-phase mixture powder-melt-solid has been consistently derived on basis of iso-stress homogenization and energy minimization. In this interesting contribution, the phase fractions for powder, melt and solid are treated as (independent) internal variables with associated evolution equations and compatibility (inequality) constraints.

The present work builds upon the purely thermal model developed in our previous contribution~\cite{Proell2020}.
The stress state of the three-phase mixture powder-melt-solid is consistently described on basis of an iso-strain homogenization across these phases.
As a main contribution we motivate and derive a constitutive model which ensures a stress-free state in newly solidified material by means of a reference strain term, formulated in rate form.
Compared to existing approaches, the proposed scheme allows to consistently account for this stress-free solidification start while avoiding spatial and temporal discontinuities (jumps) in the stress-field, as typically resulting from (instantaneous) stress resetting procedures. Moreover, it results in a simple computational model that does not require any additional internal variables or constraint equations, and thus, can easily be integrated in existing material libraries of standard FEM codes.

The second aspect, that is addressed in a novel way in this contribution, is the problem of growing, complex geometries. In this context, various approaches have been utilized to model newly added powder layers, e.g. so-called \textit{quiet-element} methods \cite{Denlinger2014}, \textit{element-birth} methods \cite{Neiva2019} or combinations thereof~\cite{Michaleris2014}, most of which use layerwise conforming FE meshes. 
    Only very few approaches, e.g., based on the immersed boundary method \cite{Carraturo2020}, can be found that allow for more flexible FE discretizations that do not need to conform with the part shape

In this contribution we propose to apply new powder layers with dual mortar mesh tying schemes, which enable efficient condensation of constraint equations from the global system of equations. This procedure offers superior flexibility with regard to spatial discretization by allowing for non-conforming meshes between subsequent layers. As compared to collocation-type coupling schemes, e.g., based on hanging nodes, mortar methods enforce the coupling conditions between non-matching meshes in an $L^2$-optimal manner. Non-conforming interface discretizations between powder layers can significantly simplify mesh generation for complex part geometries, especially, when the cross section of the part changes rapidly. We demonstrate this aspect and the general capabilities and robustness of the mortar approach based on a larger three-dimensional example.

Finally, it should be noted that this work is mainly concerned with the numerical modeling of PBFAM, a family of processes with selective laser melting (SLM) as the most prominent member. However, the methodology presented herein may also be applied to the modeling of directed energy deposition or wire-feed AM processes.

The remainder of this article is structured as follows: Section~\ref{sec:mathematical} presents the underlying mathematical model of thermomechanics. More specifically, Section~\ref{sec:temp_phase_params} reviews the modeling of temperature- and phase-dependent thermal material parameters and Section~\ref{sec:mechanical_material_law} presents the newly proposed solid material model. Its numerical treatment is detailed in Section~\ref{sec:numerical_procedure} along with the general numerical solution procedure. The model is verified by elementary test cases allowing for analytic solutions in Section~\ref{sec:validation_examples}. Section \ref{sec:meshtying} introduces the mortar mesh tying framework for adding new powder layers. Its capabilities are demonstrated in a variety of three-dimensional examples in Section \ref{sec:three_dim_examples}. Finally, Section~\ref{sec:conclusion} concludes the article with a summary of the most important results and observations.

\section{Mathematical problem statement}
\label{sec:mathematical}

The considered thermomechanical problem consists of the dynamic heat equation coupled with the static balance of linear momentum:
\begin{align}
    \label{eq:heat_equation}
    c(T)\, \dot{T} +\nabla \cdot \vec{q} &= \hat{r} \\
    \label{eq:balance_linear_momentum}
    \nabla \cdot \vec{\sigma} &= \vec{0}
\end{align}
with the primary variables temperature $T$ and displacement $\vec{u}$.
The magnitudes of displacements and rotations as arising from typical PBFAM processing conditions can be assumed as small, hence the problem is commonly modeled within the theory of geometrically linear continuum mechanics where the (engineering) strain tensor is defined by:
\begin{align}
    \label{eq:kinematics_epsilon}
    \vec{\varepsilon} = \frac{1}{2}\left(\nabla \vec{u} + (\nabla \vec{u})^{T}\right)
\end{align}
The coupling between the two equations \eqref{eq:heat_equation} and \eqref{eq:balance_linear_momentum} is for now hidden in the structural material law $\vec{\sigma} = \vec{\sigma}(\vec{\varepsilon}(\vec{u}), T)$ which will be discussed in detail in Section~\ref{sec:mechanical_material_law}. The heat flux $\vec{q}$ is specified by Fourier's law of heat conduction,
\begin{align}
    \vec{q} = -k(T) \nabla T.
\end{align}
The material parameters appearing in the thermal problem, namely volumetric heat capacity $c$ and heat conductivity $k$, will in general depend on the temperature and phase. Their modeling is discussed in detail in the authors' publication~\cite{Proell2020} and is briefly reviewed in Section~\ref{sec:temp_phase_params}. The source term $\hat{r}$ is used to model the incident laser beam power based on~\cite{Gusarov2009}.
\begin{remark}%
    Technically, the heat equation~\eqref{eq:heat_equation}, which is derived from energy conservation, can contain coupled mechanical terms. The present, geometrically linear problem formulation without these coupling terms results in a one-way coupling, i.e., the temperature field influences the structural field, but not vice versa. This assumption is ubiquitous in the macroscale PBFAM simulation literature~\cite{Bruna-Rosso2020, Denlinger2015,Hodge2014, Hodge2016} and seems justified as strain-rate dependent heating effects are not relevant for a quasi-static solid mechanics problems. A model that considers the two-way coupled thermomechanical problem can be found in~\cite{Noll2020}.
\end{remark}%
The initial boundary value problem is completed by initial conditions for the temperature field and Dirichlet and Neumann boundary conditions for the thermal and structural problem:
\begin{align}
    T &= T_0,\quad &&\text{in } \Omega\text{ for }t=0,\\
    T &= \hat{T},\quad &&\text{on } \Gamma_T,\\
    \vec{q}\cdot\vec{n} &= \hat{q}, \quad &&\text{on } \Gamma_{\vec{q}},\\
    \vec{u} &= \hat{\vec{u}}, \quad &&\text{on } \Gamma_{\vec{u}},\\
    \vec{\sigma}\cdot\vec{n} &= \hat{\vec{t}}, \quad &&\text{on } \Gamma_{\vec{\sigma}}
\end{align}
where $\Omega$ is the problem domain, $\Gamma_T$ the Dirichlet and $\Gamma_{\vec{q}}$ the Neumann boundary of the thermal problem, $\Gamma_{\vec{u}}$ the Dirichlet and $\Gamma_{\vec{\sigma}}$ the Neumann boundary of the solid problem and quantities $\hat{(\cdot)}$ are prescribed values on the respective boundaries.  No initial conditions are required for the quasi-static balance of linear momentum~\eqref{eq:balance_linear_momentum}. 

In the present work an apparent capacity method accounts for the effects of latent heat. Essentially, this method modifies the heat capacity $c$, details on the derivation can again be found in~\cite{Proell2020}.

\begin{remark}[Static vs. dynamic solid mechanics problem]
    The balance of linear momentum can either be treated as a static or dynamic problem. The PBFAM process can be assumed as quasi-static as no high accelerations takes place (in the solid phase) and inertia effects are thus negligible.
Consequently, most of the literature focuses on the static structural problem \cite{Denlinger2015,Denlinger2014,Hussein2013,Parry2016}. Note that history-dependent, potentially irreversible phenomena, which play an important role for the considered class of phase change processes, are still captured by means of history information in the thermomechanical constitutive equations.
\end{remark}

\subsection{Temperature- and phase-dependent parameters}
\label{sec:temp_phase_params}
This section briefly summarizes the modeling of the three different phases powder, melt and solid. For a more detailed motivation and derivation, the reader is referred to~\cite{Proell2020}. The commonly used liquid fraction $g$ is introduced as
\begin{align}
    \label{eq:liquid_fraction}
    g(T) = \begin{cases}
        0, & T < T_s\\
        \frac{T-T_s}{T_l-T_s}, &T_s \leq T \leq T_l\\
        1, &T > T_l
    \end{cases}
\end{align}
where $T_s$ and $T_l$ represent the solidus and liquidus temperature. The irreversibility of the powder-to-melt transition is captured via the consolidated fraction
\begin{align}
    \label{eq:consolidated_fraction}
    r_c(t) = \begin{cases}
        1, & \text{if } r_c(0)=1 \text{ (i.e., initially consolidated)}\\
        \underset{\tilde{t}\leq t}{\max}\, g(T(\tilde{t})), & \text{if } r_c(0)=0 \text{ (i.e., initially powder)}\\
    \end{cases}
\end{align}
The resulting fractions of powder ($p$), melt ($m$) and solid ($s$) are computed as
\begin{align}
    r_p &= 1 - r_c,\\
    r_m &= g,\\
    r_s &= r_c -g.
\end{align}
These phase fractions can be used to interpolate arbitrary material parameters:
\begin{align}
    \label{eq:material_parameter_interp}
    f_\text{interp} = r_p(T) f_p(T) + r_m(T) f_m(T) + r_s(T) f_s(T),
\end{align}
where $f_\text{interp}$ is the interpolated parameter and $f_p$, $f_s$ and $f_m$ are the single phase parameters. This technique is applied to the thermal conductivity $k$ and the heat capacity $c$. For the mechanical material properties we refer to the next section.

\subsection{Mechanical constitutive law}
\label{sec:mechanical_material_law}
\subsubsection{Mathematical formulation}
\label{sec:material_mathematical}
An iso-strain homogenization (also known as Voigt-type homogenization) assumes that the strain in all phases is identical. Accordingly, the stress of the mixture is given by a weighted sum of the individual contributions,
a procedure that is in fact similar to the interpolation scheme \eqref{eq:material_parameter_interp}:
\begin{align}
    \label{eq:stress_weighted_sum}
    \vec{\sigma} = \sum_i r_i \vec{\sigma}_i \quad \text{with} \quad i \in \lbrace p,m,s\rbrace.
\end{align}
Based on the iso-strain assumption, the total kinematic strain \eqref{eq:kinematics_epsilon} is equal for all phases. For each single phase $i \in {p,m,s}$, it can be additively split according to
\begin{align}
    \label{eq:additive_split}
    \vec{\varepsilon}_i = \vec{\varepsilon} = \vec{\varepsilon}_{\sigma,i} + \vec{\varepsilon}_{p,i} + \vec{\varepsilon}_{T,i} + \vec{\varepsilon}_{\text{ref},i},
\end{align}
although not all terms will be utilized for each phase. The first term on the right-hand side of \eqref{eq:additive_split} is the elastic strain $\vec{\varepsilon}_{\sigma,i}$ which induces a stress $\vec{\sigma}_i$ in each phase according to a linear hyper-elastic material
\begin{align}
    \label{eq:stress_per_phase}
\vec{\sigma}_i = \ctensor_i : \veps_{\sigma,i}, 
\end{align}
where $\ctensor_i$ is the fourth-order constitutive tensor
\begin{align}
    \ctensor_i = \lambda_i\, \delta_{ab}\delta_{cd} + \mu_i (\delta_{ac}\delta_{bd} + \delta_{ad}\delta_{bc}), \quad \lambda_i = \frac{E_i\nu}{(1+\nu)(1-2\nu)},\quad \mu_i=\frac{E_i}{2(1+\nu)}.
\end{align}
The artificial Young's modulus in powder, $E_p$, and melt, $E_m$, will be chosen orders of magnitude below the physically consistent value of the solid, $E_s$. The Poisson's ratio is assumed to be the same in all phases.

The remaining terms in~\eqref{eq:additive_split} are inelastic contributions which are considered in more detail in the following. The plastic strains $\vec{\varepsilon}_{p,i}$, which are only relevant in the solid phase, could be calculated with standard approaches, e.g., an incremental problem formulation in combination with a return mapping algorithm. For simplicity, however, plastic strains will not be considered in the numerical examples in this work ($\veps_p = 0$).  The strains due to thermal expansion $\veps_T$ are assumed equal in all phases and read
\begin{align}
    \vec{\varepsilon}_{T,i} = \vec{\varepsilon}_{T}  = \vec{I}\int_{T_\text{ref}}^T\alpha_{T}\,\dd T = \alpha_{T}(T-T_\text{ref}) \vec{I},
\end{align}
where $\alpha_{T}$ is the (constant) coefficient of thermal expansion and $T_\text{ref}$ is a reference temperature. Finally, the following reference strain $\veps_{\text{ref},s} =: \veps_{\text{ref}}$, which is only relevant for the solid phase, i.e., $\veps_{\text{ref},p} = \veps_{\text{ref},m} = 0$ , is proposed in rate form:
\begin{align}
    \label{eq:epsilon_ref_rate_based}
    \veps_\text{ref} = \frac{1}{r_s}\hat{\vec{\varepsilon}}_\text{ref}, \quad \text{with} \quad \dot{\hat{\vec{\varepsilon}}}_\text{ref} = \begin{cases}
    (\veps - \veps_p -\veps_T)\, \dot{r}_s, &\text{if } \dot{r}_s > 0\\
    \hat{\veps}_\text{ref}\, \frac{\dot{r}_s}{r_s}, &\text{if } \dot{r}_s < 0\\
    0, & \text{otherwise}
\end{cases}, \quad \hat{\vec{\varepsilon}}_\text{ref}(0) = 0,
\end{align}
where $\hat{\vec{\varepsilon}}_\text{ref}$ represents an accumulation of reference strain contributions weighted by solid fractions, which is used as an intermediate variable. The first case in \eqref{eq:epsilon_ref_rate_based} refers to a solidifying material point ($\dot{r}_s > 0$) and is motivated by physics as discussed in the next section, while the second case, for a melting material point ($\dot{r}_s < 0$), ensures that the reference strain $\vepsref$ does not change during melting, i.e., $\dot{\veps}_\text{ref} = \frac{1}{r_s} \dot{\hat{\veps}}_\text{ref} - \frac{1}{r_s^2}\dot{r}_s\hat{\veps}_\text{ref} = 0$ given $\dot{r}_s < 0$. This case is necessary for a consistent notation but, as we will see later, can be circumvented in practice. Note, how the rate formulation in~\eqref{eq:epsilon_ref_rate_based} causes a continuous change in the stress over the phase change interval $[T_s; T_l]$, which is beneficial for a numerical solution, in contrast to existing approaches with an instantaneous reset of stresses at melting temperature.

For completeness, all introduced strain contributions can be inserted into~\eqref{eq:stress_weighted_sum}, which after some rearrangement yields the following total stress of the phase mixture:
\begin{align}
    \label{eq:stress_total}
\vec{\sigma} 
&= (r_p\ctensor_p + r_m\ctensor_m + r_s\ctensor_s) : (\veps - \alpha_T(T-T_\text{ref})\vec{I}) - r_s\ctensor_s:\vepsref
\end{align}
The pre-factor of the first term in \eqref{eq:stress_total} is equivalent to an average of the single phase material parameters weighted with the phase fractions $r_i$ similar to \eqref{eq:material_parameter_interp}.

\begin{remark}[Modeling assumptions]
    One of the main assumptions underlying the present and most existing thermo-mechanical PBFAM models is that mechanical stresses in the (open-surface) powder and melt phase domains are negligible. This behavior is approximated by applying a simple elastic constitutive law to these phases, with stiffness parameters that are considerably lower as compared to the solid phase, i.e., $E_p , E_m  \ll E s$ . In practice, this approximation turns out to result in moderate, i.e., limited, strains, since the deformation of these powder and melt domains is mostly kinematically controlled by the motion of the significantly stiffer solid phase domains, thus yielding only small stress contributions as desired. Moreover, as compared to approaches exactly satisfying the zero-stress assumption in powder and melt, no additional means are required for tracking and discretization of sharp interfaces inside elements. Note, the assumption that thermal strains exist also in the powder and melt phase, and are equal to thermal strains in the solid phase, has been made for simplicity here. This assumption is neither necessary nor has it a significant influence on the resulting residual stresses due to the low stiffness of these phases and the definition of reference strains \eqref{eq:epsilon_ref_rate_based}, which ensures that newly created solid material is stress-free.  Further, we assume that from the beginning powder already has the volume and density it would have after consolidation, which has to be accounted for by defining correspondingly decreased layer thicknesses. Modeling solidification shrinkage, i.e., a density increase when powder consolidates, is deemed unnecessary due to the free surface at the top of the currently processed power layer, which allows for (approximately) stress-free consolidation and shrinkage in thickness direction when the powder melts.

    The irreversibility of phase change and the reference strain \eqref{eq:epsilon_ref_rate_based} make the material behavior non-conservative.
    While the proposed approach is very general and can be combined with arbitrary (standard) solid material laws, e.g., elasto-plasticity, we purposefully restrict ourselves to purely elastic solid material behavior in the studied examples. In this scenario, the proposed reference strain term is the only non-conservative contribution to the overall material model, underlining that the creation of a stress-free state at solidification start is the most significant non-conservative aspect of the overall thermo-mechanical problem.
    
    In the simple case of purely elastic material behavior (i.e, stresses do not exceed the yield stress) a heating to a maximum temperature below melting temperature and subsequent cooling to the initial temperature would not lead to any residual stress. Only when the melting temperature is exceeded, the reference strain term will cause an (approximately) stress-free configuration at solidification start, and thus, a residual stress will remain after cooling to the initial temperature. Therefore, we identify the reference strain contribution as the minimal necessary effect for residual stress prediction in such a simplified model.  
\end{remark}

\subsubsection{Physical motivation and discussion}
\label{sec:physical_motivation_discussion}

The reference strain \eqref{eq:epsilon_ref_rate_based} and stress \eqref{eq:stress_total} form a simple yet consistent solid constitutive law. In the following, we want to discuss some properties of the material law and their physical motivation by means of analytically tractable cases. For simplicity, the plastic strain terms are neglected from now on.

Note, that the reference strains $\vepsref$ in \eqref{eq:epsilon_ref_rate_based} only change when the solid phase fraction increases according to $\dot{r}_s > 0$, i.e., for temperatures $T \in \lbrack T_s ; T_ l \rbrack$ in the phase change interval and negative temperature rates $\dot{T}  < 0$. An elastic constitutive law with low stiffness values \mbox{(i.e., $E_p , E_m  \ll E s$)}  as applied to powder and melt leads to small stresses yet considerable total strains in these phases. In this context, the reference strains according to \eqref{eq:epsilon_ref_rate_based} ensure that these strains do not translate into stresses during solidification.
For the special case that kinematic $\veps$ and thermal strains $\veps_T$  (as well as plastic  strains $\veps_p$) are constant during solidification, which approximately holds if the phase change interval $T_l - T_s$ is sufficiently small, it can easily be verified from \eqref{eq:additive_split} and \eqref{eq:epsilon_ref_rate_based} that the elastic strain, and thus due to \eqref{eq:stress_per_phase} the resulting stresses, in the evolving solid phase vanish. This fact corresponds to the physical intuition that newly formed solid should lose all history information and exhibit a new stress-free configuration when solidification starts.

\begin{figure}
    \centering
    \includegraphics[width=.95\linewidth]{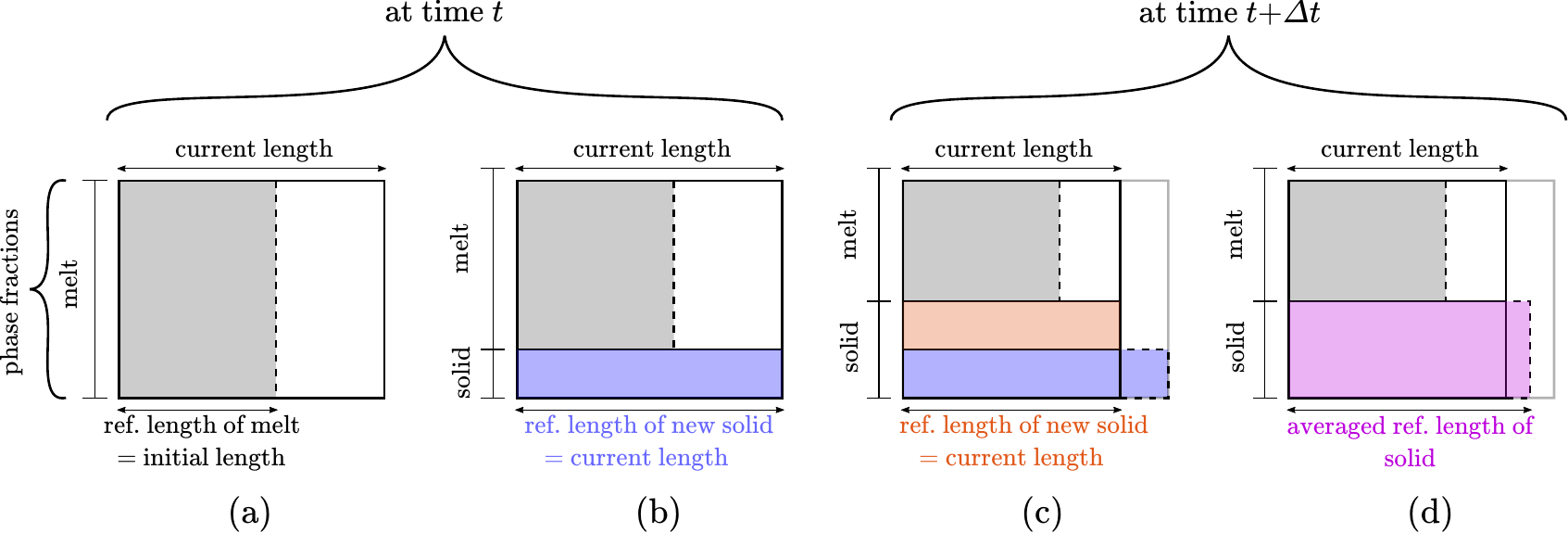}
    \caption{Effect of the reference strain contribution during solidification in a one-dimensional setting. Different phase fractions are distinguished in vertical direction together with their respective reference length, visualized in horizontal direction. At time $t$, starting from a fully molten phase (a), a fraction of material solidifies  (b) and due to the calculation of the reference strain effectively takes on a new reference configuration equal to the current configuration (blue). At time $t+\Delta t$, the newly solidified fraction takes on the (now changed) current configuration (red) as reference configuration (c), such that the effective reference strain (which in fact is calculated in the proposed model) in the (total) solid phase can be interpreted as weighted average (purple) of the contributions from individual solid phase fraction increments (d).}
    \label{fig:solidifcation_ref_strain}
\end{figure}

From a stress homogenization point of view, the model assumptions underlying the reference strain formulation state that for each solidifying fraction of material, the reference strain contribution effectively creates its new, stress-free reference configuration, from which strains are calculated. This is illustrated in Figure~\ref{fig:solidifcation_ref_strain}.

Two more involved examples, which illustrate the evolution of the reference strain over multiple repeated melt and solidification cycles, can be found in Appendix~\ref{sec:appendix_ref_strain_behavior}.

\section{Numerical solution procedure}
\label{sec:numerical_procedure}
Spatial discretization of the partial differential equation is based on the FEM. The required weak form of the thermomechanical problem \eqref{eq:heat_equation} and \eqref{eq:balance_linear_momentum} is obtained via multiplication with test functions $\delta \vec{u}$ and $\delta T$ and integration by parts, resulting in
\begin{align}
\label{eq:heat_equation_weak}
\int_\Omega \delta T\,c(T)\dot{T}\, \dd\Omega - \int_\Omega \nabla\delta T\cdot\vec{q}\,\dd\Omega + \int_{\Gamma_{\vec{q}}}\delta T\,\hat{q}\,\dd\Gamma-\int_\Omega\delta T\, \hat{r}\,\dd\Omega &= 0,\\
\label{eq:balance_linear_momentum_weak}
\int_\Omega \delta \vec{\varepsilon} : \vec{\sigma}\,\dd\Omega - \int_{\Gamma_{\vec{\sigma}}} \delta \vec{u} \cdot \hat{\vec{t}} \, \dd \Gamma &= 0
\end{align}
where the boundary conditions have already been inserted. \eqref{eq:heat_equation_weak} and \eqref{eq:balance_linear_momentum_weak} are equivalent to the strong forms~\eqref{eq:heat_equation} and \eqref{eq:balance_linear_momentum} if the solution functions are chosen from the trial spaces $\mathcal{V}_{T} \!=\! \{T \!\in\! \mathcal{H}^1(\Omega): T\!=\!\hat{T} \, \text{on} \, \Gamma_T \}$ 
and $\mathcal{V}_{\vec{u}} \!=\! \{\vec{u} \!\in\! \mathcal{H}^1(\Omega): \vec{u}\!=\!\hat{\vec{u}} \, \text{on} \, \Gamma_{\vec{u}} \}$, and
the test functions are chosen from the weighting spaces \mbox{$\mathcal{W}_{\delta T} \!=\! \{\delta T \!\in\! \mathcal{H}^1(\Omega): \delta T\!=\!0 \, \text{on} \, \Gamma_T \}$} and \mbox{$\mathcal{W}_{\delta \vec{u}} \!=\! \{\delta \vec{u} \!\in\! \mathcal{H}^1(\Omega) : \delta \vec{u}\!=\!\vec{0} \, \text{on} \, \Gamma_{\vec{u}} \}$}.
Here, $\mathcal{H}^1(\Omega)$ denotes the Sobolev space of functions with square-integrable first derivatives. The solution and test functions are discretized with a Bubnov-Galerkin ansatz in space. 
\begin{align}
T(\vec{x},t) &= \sum_j N_j(\vec{x})T_j(t), \quad &\delta T(\vec{x},t) &= \sum_j N_j(\vec{x})\delta T_j(t)\\
\vec{u}(\vec{x},t) &= \sum_j N_j(\vec{x})\vec{d}_j(t), \quad &\delta \vec{u}(\vec{x},t) &= \sum_j N_j(\vec{x})\delta \vec{d}_j(t)
\end{align}
where $\vec{x}$ is the spatial position and $N_j(\vec{x})$ are the time-independent shape functions, which are the same for all solution and test functions. $T_j$ and $\delta T_j$ are the discrete nodal temperatures and their variations, and $\vec{d}_j$ and $\delta \vec{d}_j$ are the discrete nodal displacements and their variations, all of which depend on time. Although the static balance of linear momentum \eqref{eq:balance_linear_momentum} is considered, i.e., inertia forces are neglected, the displacement field $\vec{u}(\vec{x},t)$ is depending on time via the coupling to the temperature field ${T}(\vec{x},t)$, which is a solution to the dynamic heat equation \eqref{eq:heat_equation}. 
The thermal subproblem is discretized in time with a generalized trapezoidal rule. The discrete system of equations is consistently linearized and solved with Newton's method.

\subsection{Discretization of the rate-based reference strain}
\label{sec:numerical_ref_strain}

Time integration of equation \eqref{eq:epsilon_ref_rate_based} is based on a backward Euler scheme and results in the following time-discrete form of the fraction-weighted reference strain  $\hat{\vec{\varepsilon}}_\text{ref}^{n+1}$:
\begin{align}
    \label{eq:epsilon_ref_integrated}
    \hat{\vec{\varepsilon}}_\text{ref}^{n+1} =
    \begin{cases}
    \hat{\vec{\varepsilon}}_\text{ref}^{n} + \Delta r_s^{n+1}(\veps^{n+1} -\veps_T^{n+1}) & \text{if}\ \Delta r_s^{n+1} > 0,\\
    \hat{\vec{\varepsilon}}_\text{ref}^{n} + \frac{\Delta r_s^{n+1}}{r_s^{n+1}}\hat{\veps}_\text{ref}^{n+1} & \text{if}\ \Delta r_s^{n+1} < 0,\\
    \hat{\vec{\varepsilon}}_\text{ref}^{n} & \text{otherwise},
    \end{cases}
\end{align}
which leads to the actual reference strains via relation \eqref{eq:epsilon_ref_rate_based} as $\vepsref^{n+1} = \frac{1}{r_s^{n+1}}\hat{\veps}_\text{ref}^{n+1}$.
Thus, the time-discrete total reference strain $\vepsref^{n+1}$ may also be obtained directly, viz.
\begin{align}
    \label{eq:epsilon_ref_total_recursive}
    \vepsref^{n+1} =
    \begin{cases}
    \frac{1}{r_s^{n+1}}\left(\vepsref^{n} r_s^n + \Delta r_s^{n+1}(\veps^{n+1} -\veps_T^{n+1}) \right),& \text{if}\ \Delta r_s^{n+1} > 0\\
    \vepsref^{n}, & \text{otherwise}
    \end{cases}
\end{align}
Note, that in \eqref{eq:epsilon_ref_total_recursive} the case of melting material ($\Delta r_s^{n+1} < 0$) no longer needs to be treated separately due to the construction of $\hat{\veps}_\text{ref}$ as already discussed when it was first introduced.

\subsection{Linearization of constitutive law}
The linearization of the total stress \eqref{eq:stress_total} with respect to the primary solution variables is required for the nonlinear solution procedure. As usual, the derivatives of the constitutive equation are written with respect to the kinematic strain and temperature, 
\begin{align}
    \pfrac{\vec{\sigma}}{\vec{\varepsilon}^{n+1}} &= (r_p\ctensor_p + r_m\ctensor_m + r_s\ctensor_s) - \Delta r_s^{n+1} \ctensor_s\,\heaviside{\Delta r_s^{n+1}}\\
    \label{eq:dsigma_dT}
    \pfrac{\vec{\sigma}}{T^{n+1}} &= \left(\pfrac{r_p}{T}\ctensor_p + \pfrac{r_m}{T}\ctensor_m + \pfrac{r_s}{T}\ctensor_s\right): \left(\veps^{n+1}-\alpha_T(T^{n+1}-T_0)\vec{I}\right) \nonumber\\ 
    &-(r_p\ctensor_p + r_m\ctensor_m + r_s\ctensor_s):\alpha_T\vec{I} \nonumber\\
    &-\ctensor_s:\left(\pfrac{r_s}{T}\left( \veps^{n+1} -\veps_T^{n+1} \right) - \Delta r_s^{n+1}\alpha_T\vec{I}\right)\,\heaviside{\Delta r_s^{n+1}}\nonumber\\
    & - \pfrac{r_s}{T} \ctensor_s : \vepsref^{n}\,\heaviside{-\Delta r_s^{n+1}}
\end{align}
where $\heaviside{x}$ is the Heaviside step function
\begin{align}
    \label{eq:heaviside_definition}
    \heaviside{x} = \begin{cases}
        1,\quad \text{if } x > 0\\
        0,\quad \text{otherwise}
    \end{cases}.
\end{align}
All phase fractions and their derivatives are evaluated at $T^{n+1}$. The first two terms in \eqref{eq:dsigma_dT} stem from the phase interpolation in the first term of~\eqref{eq:stress_total}. The third term represents the contribution from a solidifying increment and is efficiently derived by inserting \eqref{eq:epsilon_ref_total_recursive} into the total stress \eqref{eq:stress_total} and thereby canceling out the solid fraction $r_s$. The last term represents the effect of a melting solid. 

\section{One- and two-dimensional numerical examples}
\label{sec:validation_examples}
This section focuses on the verification of the proposed material law with a series of elementary test cases. In all of them the temperature is effectively a prescribed time-dependent function that drives the structural simulation and no heat equation needs to be solved. The investigation starts out on a one-dimensional domain which is subject to different boundary conditions and temperature profiles with increasing complexity. These findings are reported and discussed in sections \ref{sec:example_homogeneous_temperature} and \ref{sec:inhomogeneous_temperature}. Complexity is increased further by applying different temperature profiles to a two-dimensional domain in Section \ref{sec:validation_2d}.

\begin{figure}
    \centering
    \includegraphics[width=.7\linewidth]{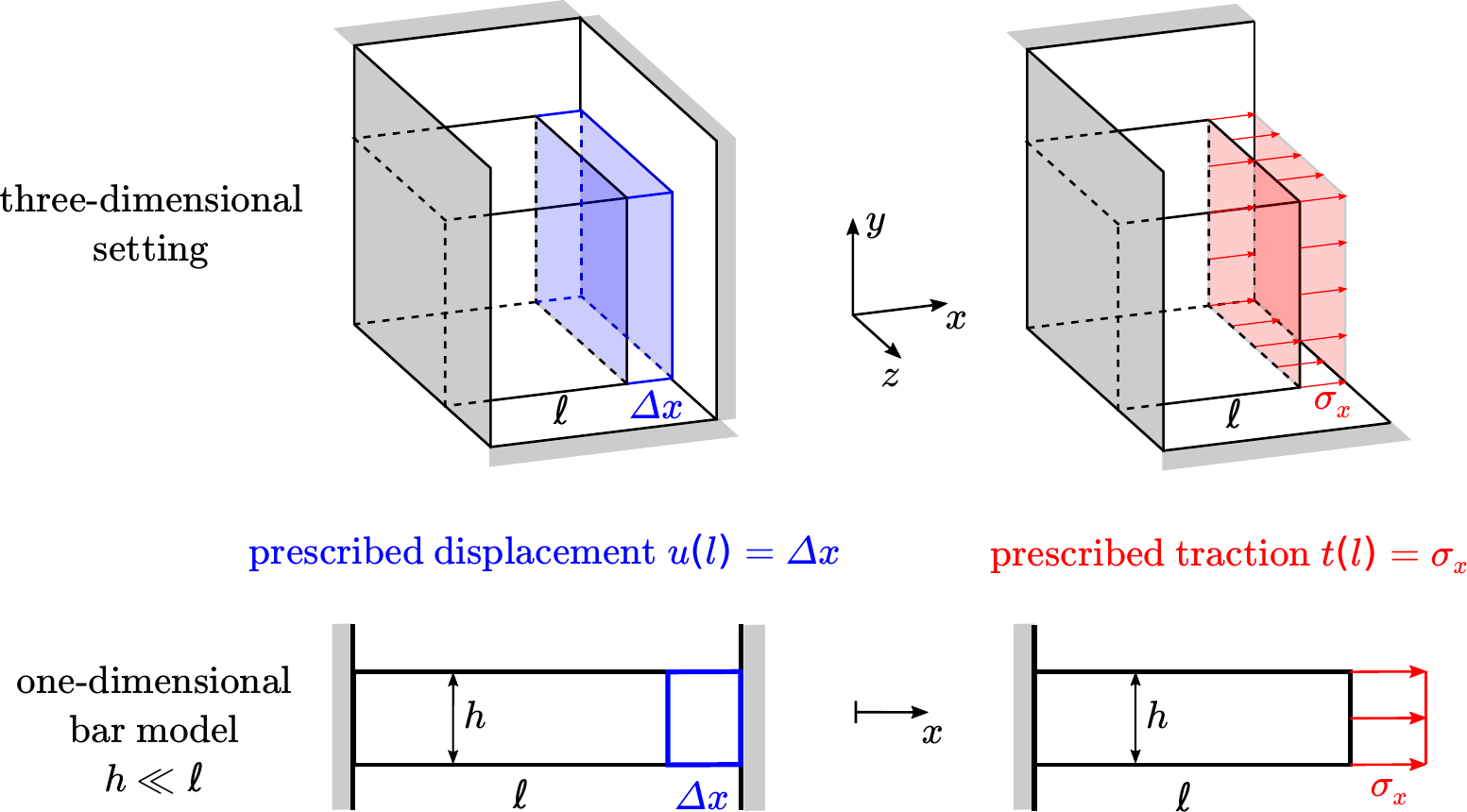}
    \caption{Relation of investigated one-dimensional {verification} examples to their respective three-dimensional solid mechanics problem.}
    \label{fig:relation_1d_3d}
\end{figure}

\begin{table}
    \centering
    \caption{Material parameters for one-dimensional simulations. Thermal properties (e.g. thermal conductivity) are not necessary for these simulations}
    \label{tab:params_1d_validation}
    \begin{tabular}{llll}
        \toprule
        Parameter & Description & Value & Unit\\
        \midrule
        $T_s$ & Solidus temperature & 1900 & \si{\celsius}\\
        $T_l$ & Liquidus temperature & 2100 & \si{\celsius}\\
        $T_0$ & Reference temperature & 0 & \si{\celsius}\\
        $E_s$ & Young's modulus in solid & \num{1} & \si{\giga\pascal} \\
        $E_p$ & Young's modulus in powder & \num{10} & \si{\mega\pascal}\\
        $E_m$ & Young's modulus in melt & \num{10} & \si{\mega\pascal}\\
        $\alpha_T$ & Coefficient of thermal expansion & \num{1e-6} & \si{\per\kelvin}\\
        \bottomrule
    \end{tabular}
\end{table}

\subsection{One-dimensional domain: homogeneous temperature load}
\label{sec:example_homogeneous_temperature}
The first series of examples examines a one-dimensional bar of length $l=1 \,\si{\milli\metre}$, which is subject to a homogeneous temperature load, i.e., the temperature evolution is only a function of time but spatially constant, leading to melting and solidification of the material, possibly multiple times. Figure~\ref{fig:relation_1d_3d} illustrates the two considered types of boundary conditions and how the respective one-dimensional problem relates to a three-dimensional solid mechanics problem. The behavior of the bar is equivalent to the behavior of the depicted cube in $x$-direction. The bar is either constrained by a Dirichlet-type condition (Figure~\ref{fig:relation_1d_3d} left) prescribing the deformation or a Neumann-type condition (Figure~\ref{fig:relation_1d_3d} right) prescribing the traction. The prescribed values can be zero (homogeneous boundary conditions) or non-zero (inhomogeneous boundary conditions). The initial material phase is either solid or powder and all relevant material parameters are listed in Table~\ref{tab:params_1d_validation}.

Three representative examples are investigated: first, a full melt of the material, and second, a repeated partial melt followed by a full melt. Both scenarios use homogeneous boundary conditions. For completeness, the partial melt scenario is repeated with inhomogeneous boundary conditions as a third example. The numerical results for all examples are compared to an analytical reference solution, which assumes isothermal melting ($T_s=T_l=T_m$) and a zero stiffness in powder and melt phase.

\begin{figure}[bt]
    \centering
    \includegraphics[width=.95\linewidth]{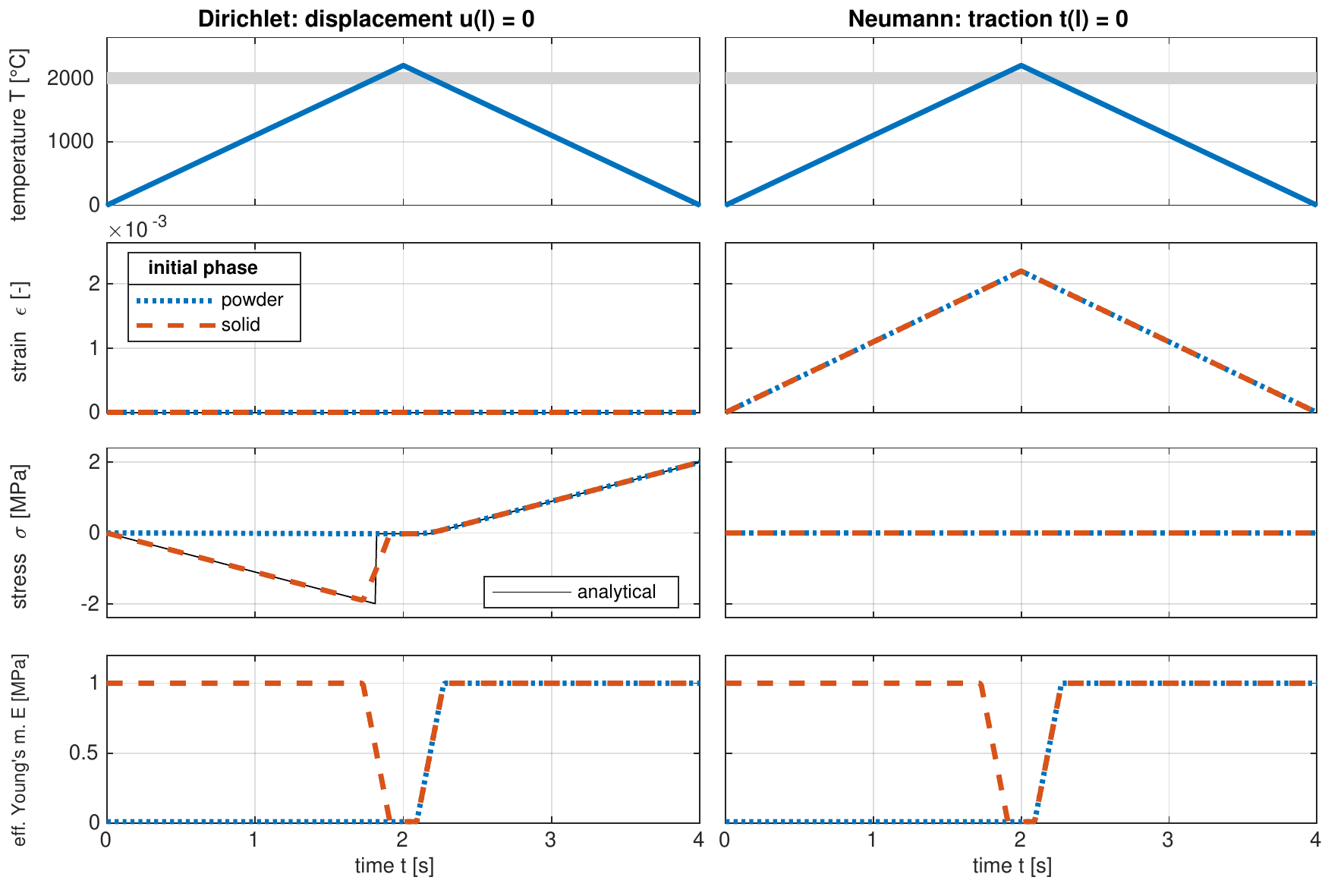}
    \caption{Stress and strain over time for the Dirichlet and Neumann scenario with initially powder or solid material induced by a full melt. Simulation is driven by a homogeneous temperature load with a single peak $T_\text{peak}>T_l$. Phase change interval in gray.}
    \label{fig:full_melt_constr_unconstr}
\end{figure}

\subsubsection{Full melt}
\label{sec:example_full_melt}
\begin{figure}[bt]
    \centering
    \includegraphics[width=.95\linewidth]{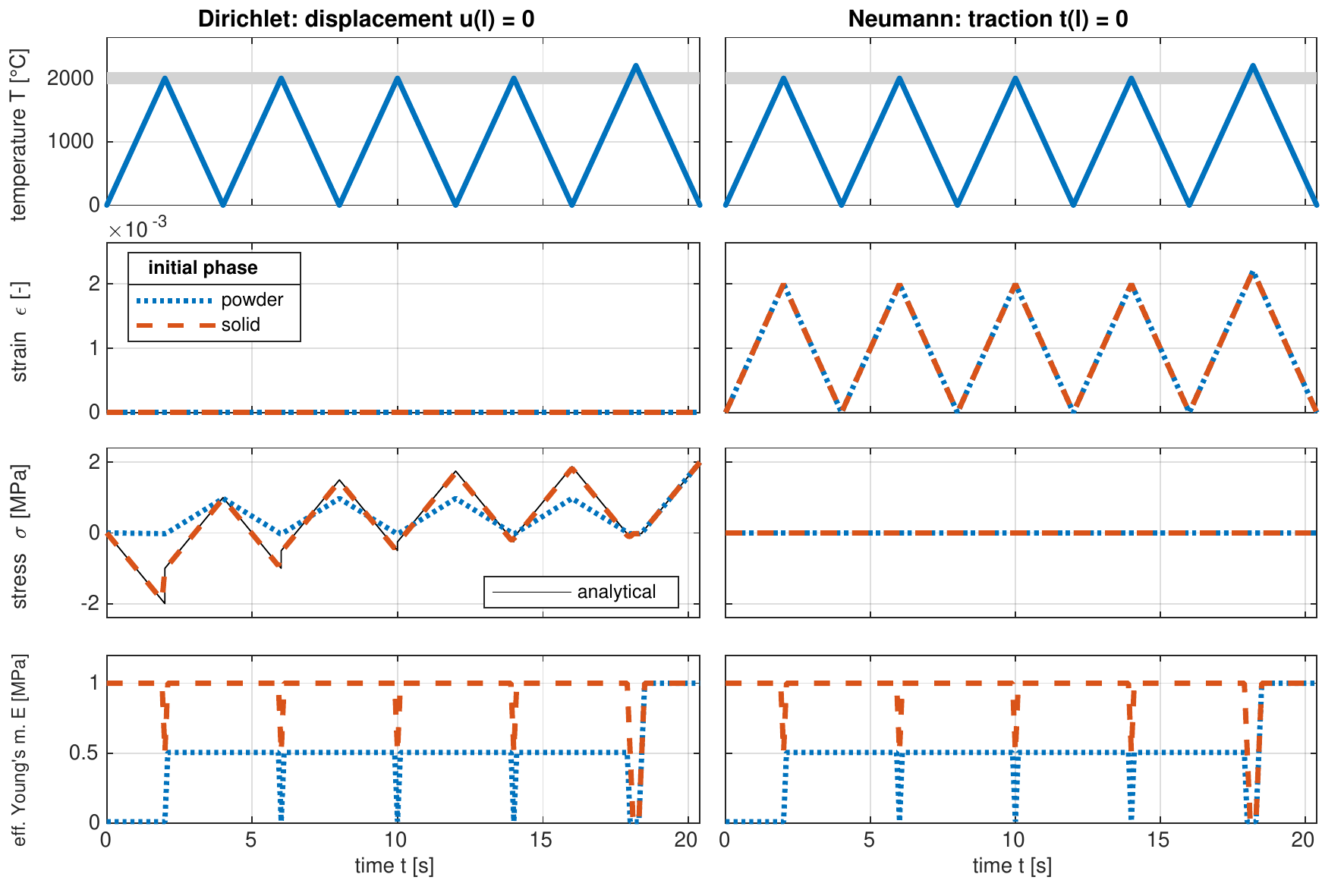}
    \caption{Stress and strain over time for the Dirichlet and Neumann scenario with initially powder or solid material induced by repeated partial melting with $T_s < \hat{T}_{1}<T_l$ and a final full melt with $\hat{T}_{2}>T_l$. Phase change interval in gray.}
    \label{fig:partial_melt_repeatedly}
\end{figure}

The material is completely molten by heating it to a peak temperature $\hat{T}=\num{2200} \,\si{\celsius} > T_l$. 
Figure~\ref{fig:full_melt_constr_unconstr} shows the temporal evolution of the (homogeneous) strain and stress state for both, Dirichlet and Neumann, scenarios as well as the evolution of the effective Young's modulus, $E = \sum r_i E_i$, of the phase mixture. While the Neumann scenario is mostly shown for completeness, the Dirichlet scenario is more interesting regarding the difference between an initially powder and solid material: before melting takes place ($t \ll 2$) the stress will stay close to a near-zero value in the case of initial powder (small stiffness) but take on significant values in the case of initial solid. When the material melts ($t\approx 2$), the stress reduces to the same small value as in the powder case. For a finite but small phase change interval, as chosen for this problem, this change happens continuously. On the other hand, the analytical reference solution for the stress, assuming isothermal melting, exhibits a discontinuity at melting temperature. The proposed model can represent this discontinuity asymptotically correct when decreasing the phase change interval. After full melting, the two stress curves for initially powder and solid material are identical since in both cases all material is now completely (re)molten and has the same reference strain. At the end of the simulation the same final stress $\sigma_\text{final} = \num{2} \,\si{\mega\pascal}$ is obtained. This value can be calculated analytically, see Appendix~\ref{sec:appendix_stress_analytical_full}.

In the scenario with a homogeneous Neumann boundary condition no stress can occur. The total strain is always equal to the thermal strain, $\varepsilon = \varepsilon_T$, and thus the contributions to the reference strain in \eqref{eq:epsilon_ref_rate_based} will always be zero. Therefore, the strain is directly proportional to the change in temperature $\Delta T = T - T_0$, as shown on the right-hand side in Figure~\ref{fig:full_melt_constr_unconstr}.

\subsubsection{Repeated partial melt followed by full melt}
\label{sec:example_partial_melt}
In a second set of examples the material is partially molten with a peak temperature $\hat{T}_1$ inside the phase change interval, viz.~$T_s < \hat{T}_1=\num{2000} \,\si{\celsius} < T_l$,  and then cooled down to $T_0$. This cycle is repeated four times and, finally, the material is fully molten with a peak temperature $\hat{T}_2=\num{2200} \,\si{\celsius} > T_l$. The resulting strain and stress evolution is shown in Figure~\ref{fig:partial_melt_repeatedly}. Once again, attention is drawn to the stress evolution in the Dirichlet case which is equal to the results from the previous section until the first peak in the temperature profile is reached. This time, the stress does not reach a near-zero value for the initially solid case because the melting stops at $\hat{T}_1 < T_l$, i.e., there remains a phase fraction of solid material at $\hat{T}_1$, which exhibits thermal stresses due to the increased temperature level as compared to the stress-free configuration at $T_0$. Subsequently, the stress rises to the same local peak value of \num{975} \si{\kilo\pascal} for both initially solid and powder material after cooling to reference temperature. This can be explained as follows: the only relevant contribution from~\eqref{eq:stress_total} is the last term containing the reference strain since the temperature is equal to the initial reference temperature and the strain is identical to zero. Although the solid fraction differs at this point (0.5 in powder case and 1 in solid case), multiplication with the reference strain \eqref{eq:epsilon_ref_rate_based} yields the same contribution $\hat{\vec{\varepsilon}}_\text{ref}$. Put differently: independent of the initial phase, the non-molten phase fraction yields a stress of zero, since $T=T_0$; the (re)molten phase fraction yields the same stress contribution, since the (re)molten fraction is equal for initially powder and solid material (same maximum temperature). The specific value of \num{975} \si{\kilo\pascal} can be calculated analytically as demonstrated in Appendix~\ref{sec:appendix_stress_analytical_partial}.

The same heating and cooling cycle is repeated three times. For an initial powder material the stress reaches a near-zero value at $\hat{T}_1$ because the solid phase (from the previous partial melt) is completely remolten at this point. For an initially solid material, the situation was already discussed in Section~\ref{sec:physical_motivation_discussion}: the material model contains no information about different solid phases but instead averages the reference contributions according to~\eqref{eq:epsilon_ref_total_recursive}.
As already outlined before, a repeated partial melt with the same peak value will lead to increasing reference strain and, therefore, also stresses (when cooled to reference temperature), a fact that is well illustrated by the simulation results in Figure~\ref{fig:partial_melt_repeatedly}. The analytical reference solution for isothermal melting of initially solid material is also included and again only differs for temperatures in the phase change interval (although not visible in Figure~\ref{fig:partial_melt_repeatedly}).

Finally, the material is fully molten with a peak temperature $\hat{T}_2=\num{2200} \,\si{\celsius} > T_l$ and cooled to $T_0$. Both an initially solid and powder material yield the same final stress value of \num{2} \si{\mega\pascal}, which is the same value as in the previous example, where the material was fully molten directly in one thermal cycle. Once more, this confirms that all history information related to stresses is erased when all solid is (re)molten.

\subsubsection{Inhomogeneous boundary conditions}
\label{sec:example_inhomogeneous_BCs}
The scenarios so far were only concerned with zero Dirichlet and Neumann boundary conditions. For completeness the partial melt case in Section~\ref{sec:example_partial_melt} is repeated with an inhomogeneous Dirichlet-type constraint $u(l) = 0.001\, \si{\milli\metre}$, which leads to a constant strain $\varepsilon_x = 0.001$ for the given homogeneous temperature load. 

\begin{figure}[btp]
    \centering
    \includegraphics[width=.8\linewidth]{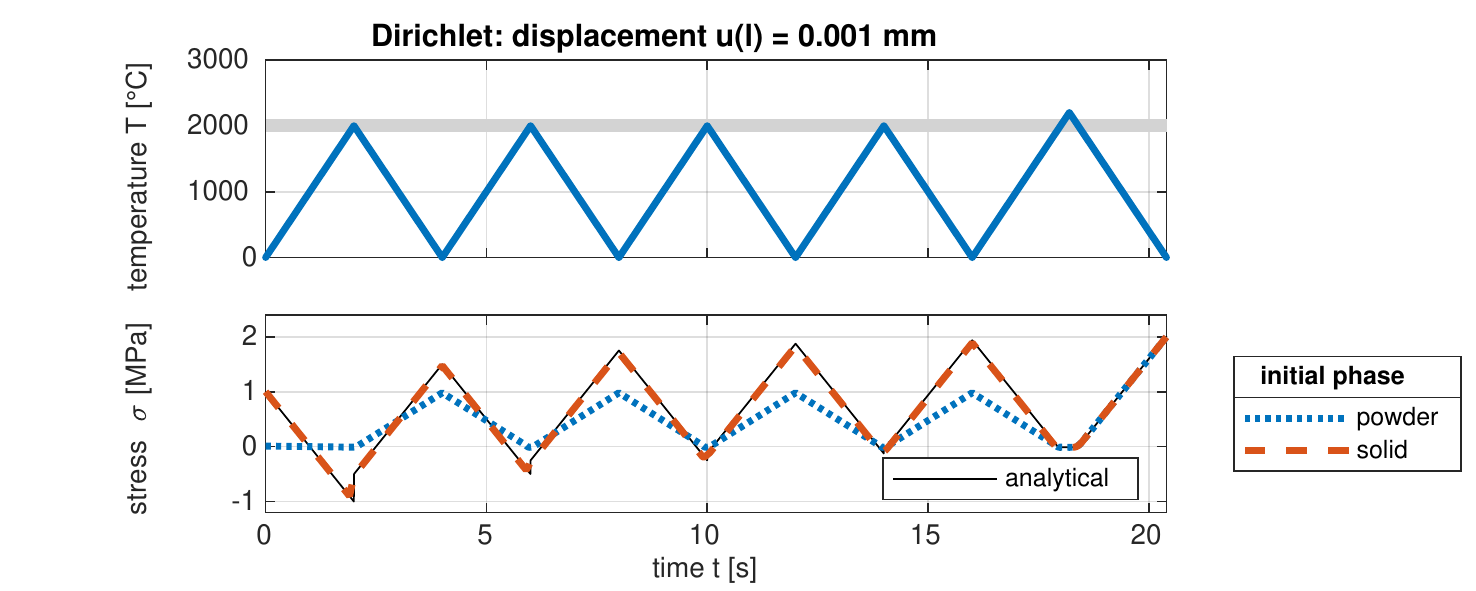}
    \caption{Stress over time induced by a repeated partial melt  and a final full melt. Inhomogeneous Dirichlet boundary conditions. Phase change interval in gray. }
    \label{fig:partial_melt_inhomogeneous_bcs}
\end{figure}

Figure~\ref{fig:partial_melt_inhomogeneous_bcs} shows the resulting residual stresses, which, for an initially solid material, differs from Figure~\ref{fig:partial_melt_repeatedly}: the fixed strain induces an initial stress that decreases during the repeated partial melting cycles and, eventually, completely vanishes after the full melting, during which all pre-existing residual stresses are relaxed. Again, the behavior of the solid after full melting is identical to the examples in Sections~\ref{sec:example_full_melt} and \ref{sec:example_partial_melt}. For an initially powder material, the results are almost unaffected by the inhomogeneous Dirichlet value, which is consistent with the basic modeling assumption that mechanical stresses resulting from a deformation of the powder phase are negligible.

\subsection{One-dimensional domain: inhomogeneous temperature load}
\label{sec:inhomogeneous_temperature}
\begin{figure}
    \centering
    \includegraphics[width=.4\linewidth]{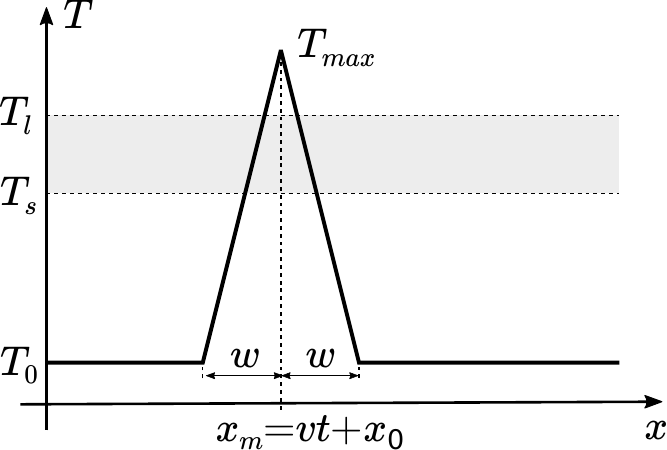}
    \caption{Schematic of temperature peak function, which will move with speed $v$ in $x$-direction.}
    \label{fig:tpeak}
\end{figure}

\begin{figure}[btp]
    \centering
    \includegraphics[width=\linewidth]{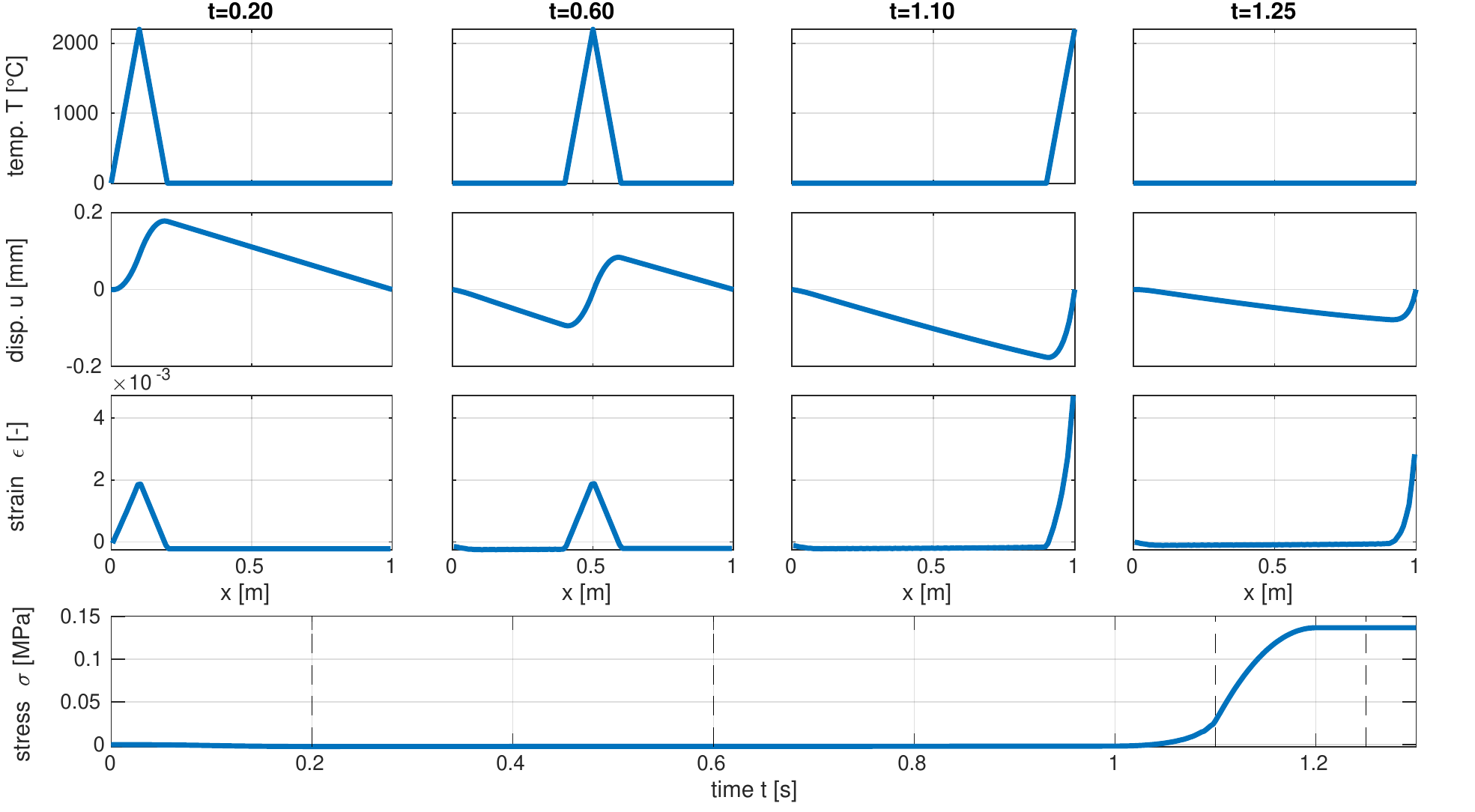}
    \caption{Temperature, displacement and strain at selected time steps (first three rows) and stress evolution over time for the inhomogeneous temperature load example. Dashed black lines in the time-stress diagram indicate the selected time steps for the upper diagrams.}
    \label{fig:inhomogeneous_temperature_peak}
\end{figure}
The next scenario focuses on a more complex, transient and inhomogeneous temperature profile and a homogeneous Dirichlet boundary condition $u(l) = 0$ (see left-hand side in Figure~\ref{fig:relation_1d_3d}). A moving temperature peak sketched in Figure~\ref{fig:tpeak}, that starts outside the initially powder domain (length $l=1\,\si{\metre}$) and moves through it to the other side, mimics the effect of a moving laser beam in PBFAM. For completeness, the mathematical form of such a temperature profile is stated as
\begin{align}
    \label{eq:1d_temperature_peak_profile}
    T(x,t) = T_0 
    + \frac{T_\text{max}-T_0}{w} \times \begin{cases}
        w-\hat{x}, &\text{if } 0\leq \hat{x} < w\\
        w+\hat{x},  &\text{if } -w< \hat{x} < 0\\
        0, &\text{otherwise}\\
    \end{cases}, \ \text{with } \hat{x} = x-vt-x_0
\end{align}
where the parameters for the present example are chosen as $T_0=0$, $T_\text{max}=2200\,\si{\celsius}$, $w=0.1\,\si{\metre}$, $x_0 = -w$ and $v=1.0\,\si{\metre\per\sec}$. With these parameters every point in the domain melts and solidifies.
Figure~\ref{fig:inhomogeneous_temperature_peak} shows the temperature profile and the resulting displacement and strain field at select time steps. The stress, which is constant over the domain, is plotted over the whole simulation time. In the beginning, the low stiffness of the remaining powder at the right side of the domain enables an almost free thermal expansion and the stress is close to zero. It increases to significant values only when the whole bar has a solid fraction $r_s >0$, i.e., when the temperature peak reaches the right end of the bar, where the displacement is fixed to zero.

\begin{figure}[btp]
    \centering
    \includegraphics[width=.46\linewidth]{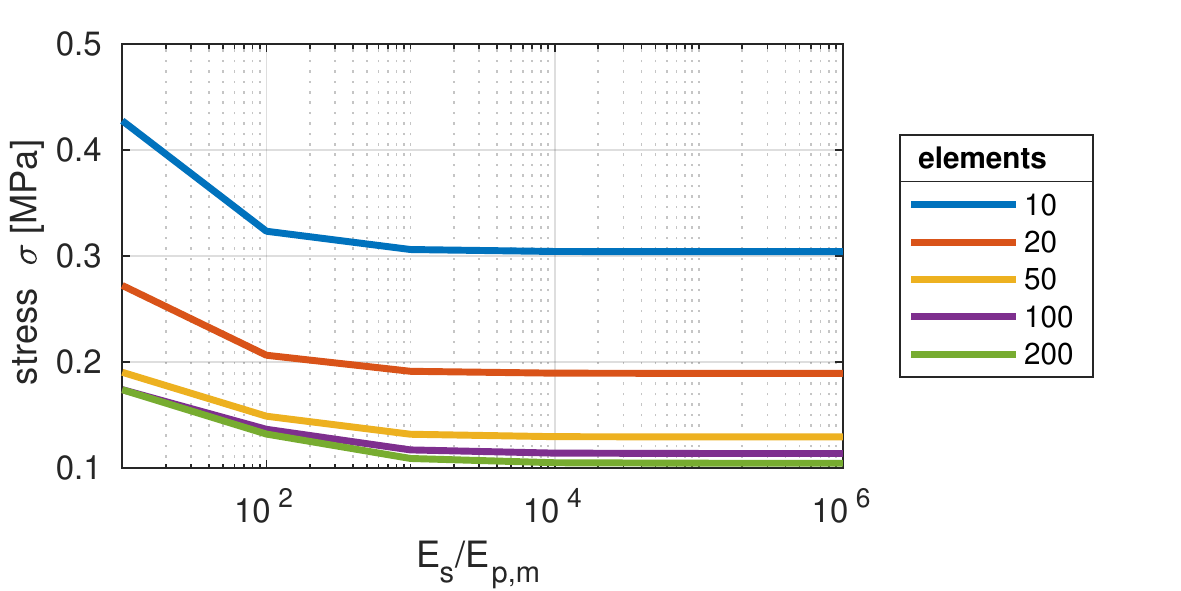}
    \includegraphics[width=.46\linewidth]{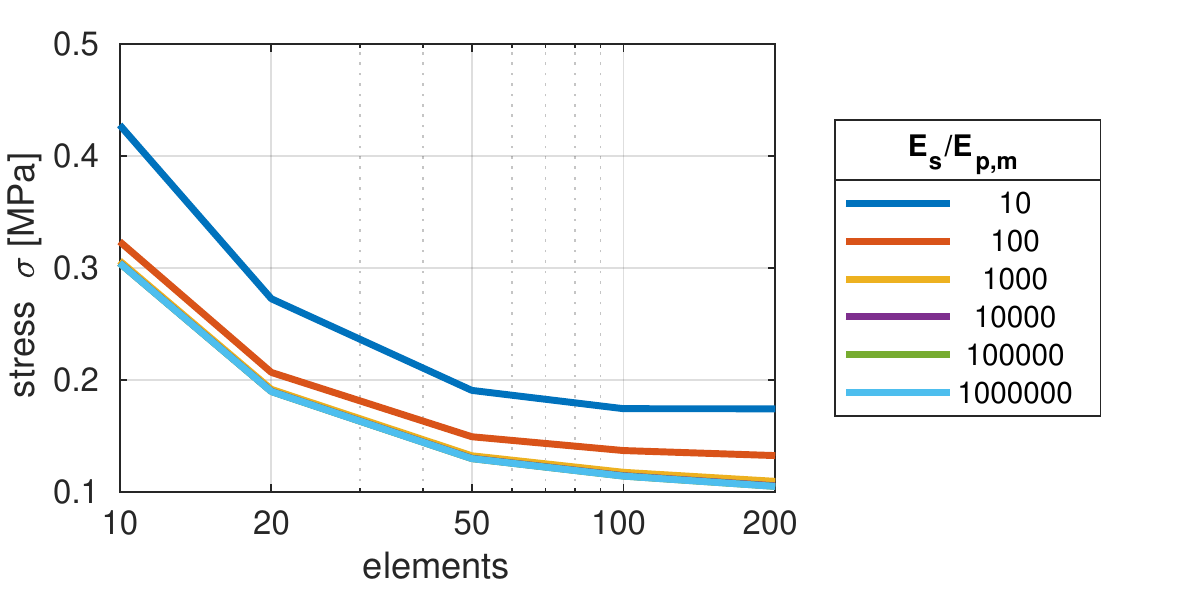}
    \caption{Convergence of the final stress in a constrained bar subject to a moving peak temperature load. Convergence is influenced by the number of (finite) elements and the stiffness ratio between solid and powder/melt. Both diagrams contain the same information in different representations. }
    \label{fig:convergence_single_slab_peak_source_constr}
\end{figure}

The current example is also used to judge the influence of the stiffness ratio $\frac{E_s}{E_{p,m}}$ between the solid and powder/melt Young's moduli by varying the (artificial) powder/melt stiffness $E_{p,m}$.
Figure~\ref{fig:convergence_single_slab_peak_source_constr} shows how the final stress in the bar (after cooldown to the reference temperature) converges with a refined spatial discretization and for different stiffness ratios. For a high stiffness ratio the limit value lies at around \num{0.1}\,\si{\mega\pascal}, which one would obtain as the analytical value  assuming the theoretical value of zero powder and melt stiffness, i.e. a stiffness ratio of infinity. A proof is sketched in Appendix \ref{appendix:analytical_1d_inhomogeneous}.

\subsection{Two-dimensional domain: interaction of boundary condition and temperature load}
\label{sec:validation_2d}

\begin{figure}[btp]
    \centering
    \includegraphics[width=.5\linewidth]{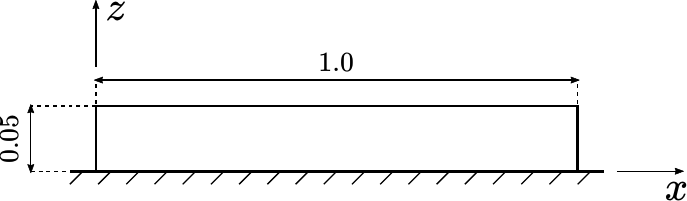}
    \caption{Geometry and boundary conditions of two-dimensional domain: bottom edge fixed; left, right and top edge free. Dimensions in \si{\milli\metre} (not to scale).}
    \label{fig:two_dimensional_geometry}
\end{figure}

\begin{figure}[btp]
    \centering
    \includegraphics[width=.95\linewidth]{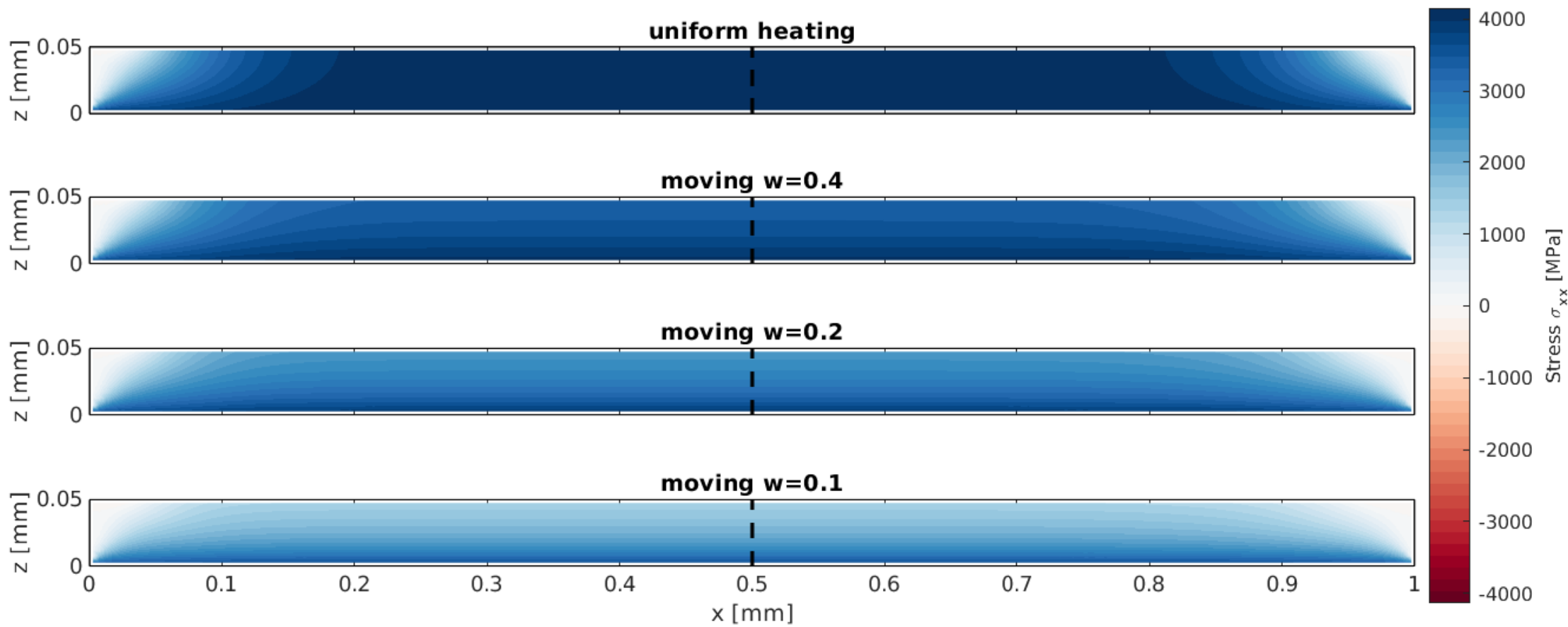}
    \caption{Two-dimensional example: residual stress distribution $\sigma_{xx}$ after subjecting domain to a uniform  (first row) or moving (second to fourth row) peak temperature profile.}
    \label{fig:2d_uniform_vs_moving_whole_domain}
\end{figure}

The final verification example investigates a two-dimensional setting under the assumption of plane-stress\footnote{A plane-strain setting, which was also simulated by the authors, would change the absolute values but not the qualitative trends presented in this section.}. As in the previous section, the geometry consists of a slender, bar-like structure of powder material which is now constrained with a homogeneous Dirichlet boundary condition on the bottom edge ($z=0$), while the left, right and top edges are free (homogeneous Neumann boundary condition), see Figure~\ref{fig:two_dimensional_geometry}. The motivation for this setup is to mimic the processing of a powder layer which is applied atop an already solid domain. For this example we use the realistic material parameters from Table~\ref{tab:params_3d}.

\begin{figure}[bt]
    \centering
    \includegraphics[width=.46\linewidth]{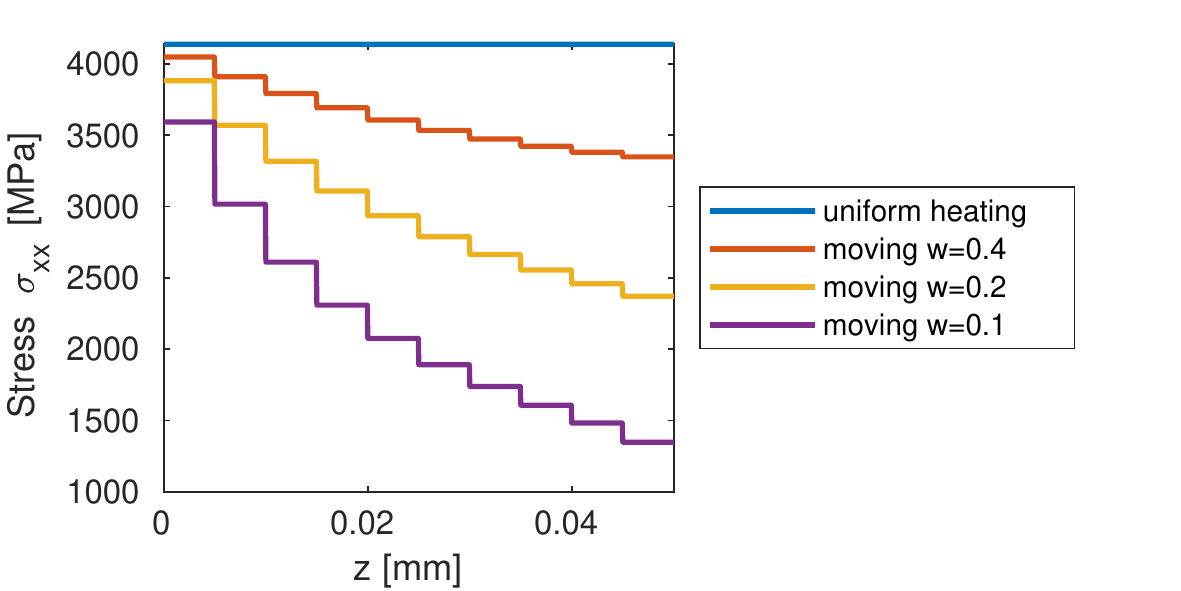}
    \includegraphics[width=.46\linewidth]{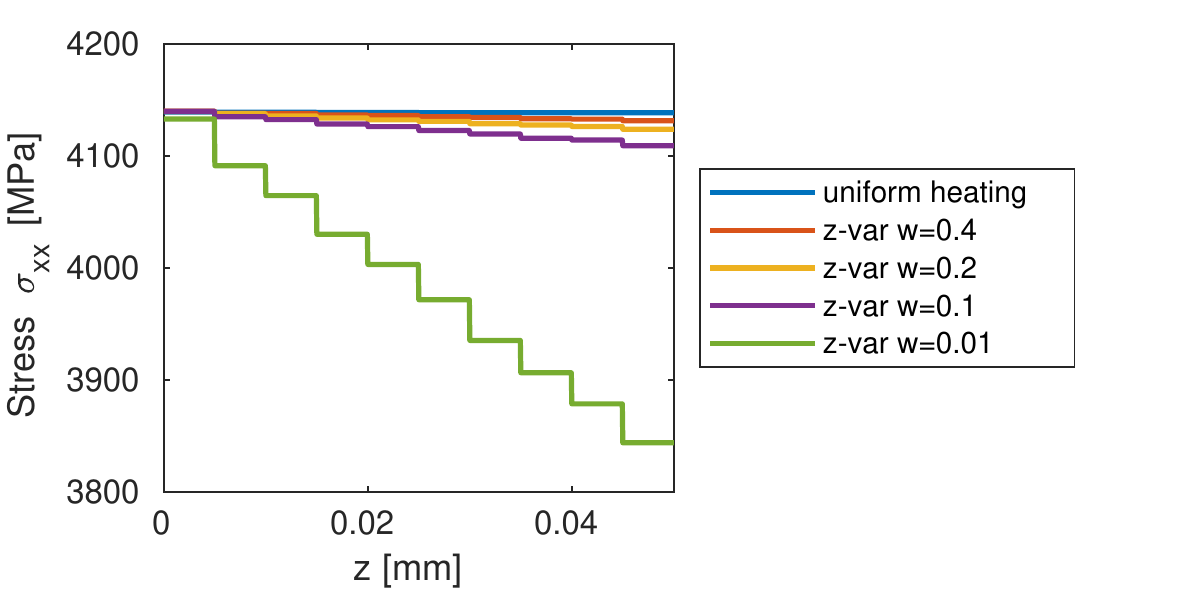}
    \caption{Two-dimensional example: final stress $\sigma_{xx}$ over height $z$ (zero is constrained bottom) at $x=\num{0.5}\,\si{\milli\metre}$ (see dashed line in Figure~\ref{fig:2d_uniform_vs_moving_whole_domain}): induced by a temperature peak moving in $x$-direction, constant in $z$-direction (left);  induced by a temperature profile varying in $z$-direction, constant in $x$-direction (right). Note the different ranges of the ordinates.}
    \label{fig:2d_zcut}
\end{figure}

As a baseline the domain is heated uniformly across the entire domain from initial temperature $T_0=303\,\si{\kelvin}$ to a maximum temperature $T_\text{max}=2000\,\si{\kelvin} > T_l$ and then cooled down to the initial temperature. This case is compared to the scenario of a moving temperature peak profile as in \eqref{eq:1d_temperature_peak_profile} with different peak widths $w \in \lbrace 0.4,0.2,0.1\rbrace\, \si{\kelvin}$ and $T_\text{max}=2000\,\si{\kelvin}$. The peak still moves only in $x$-direction with speed $v=1.0 \,\si{\milli\metre\per\second}$, i.e., the temperature is considered constant in $z$-direction, which is a rough approximation to the temperature profile encountered in a PBFAM application. Figure~\ref{fig:2d_uniform_vs_moving_whole_domain} compares the resulting residual stresses $\sigma_{xx}$ in the different scenarios: a uniform temperature profile yields the highest tensile stresses and almost no variation of the stress $\sigma_{xx}$ in $z$-direction (the ``powder layer height'') while a moving temperature peak profile leads to a significant drop-off of $\sigma_{xx}$ over the height, meaning that stresses are lower at the non-constrained upper surface.
This drop-off becomes more pronounced with a decreased peak width $w$ and can be more clearly observed in Figure~\ref{fig:2d_zcut} (left). These observations can be explained by the size of the domain which has a temperature above $T_0$ and is cooling down: if this region is small (for a small $w$), the deformation occuring during cooldown is almost unconstrained since these solid material regions with elevated temperatures are very close to the (approximately) stress-free liquid domain, thus resulting in a comparatively small tensile stress. If the cooling region is wider, large portions of the cooling solid material are far away from the stress-free solid-liquid boundary and thus are strongly constrained in their ability to deform, which consequently results in a higher tensile stress.

\begin{figure}[bt]
    \centering
    \includegraphics[width=.95\linewidth]{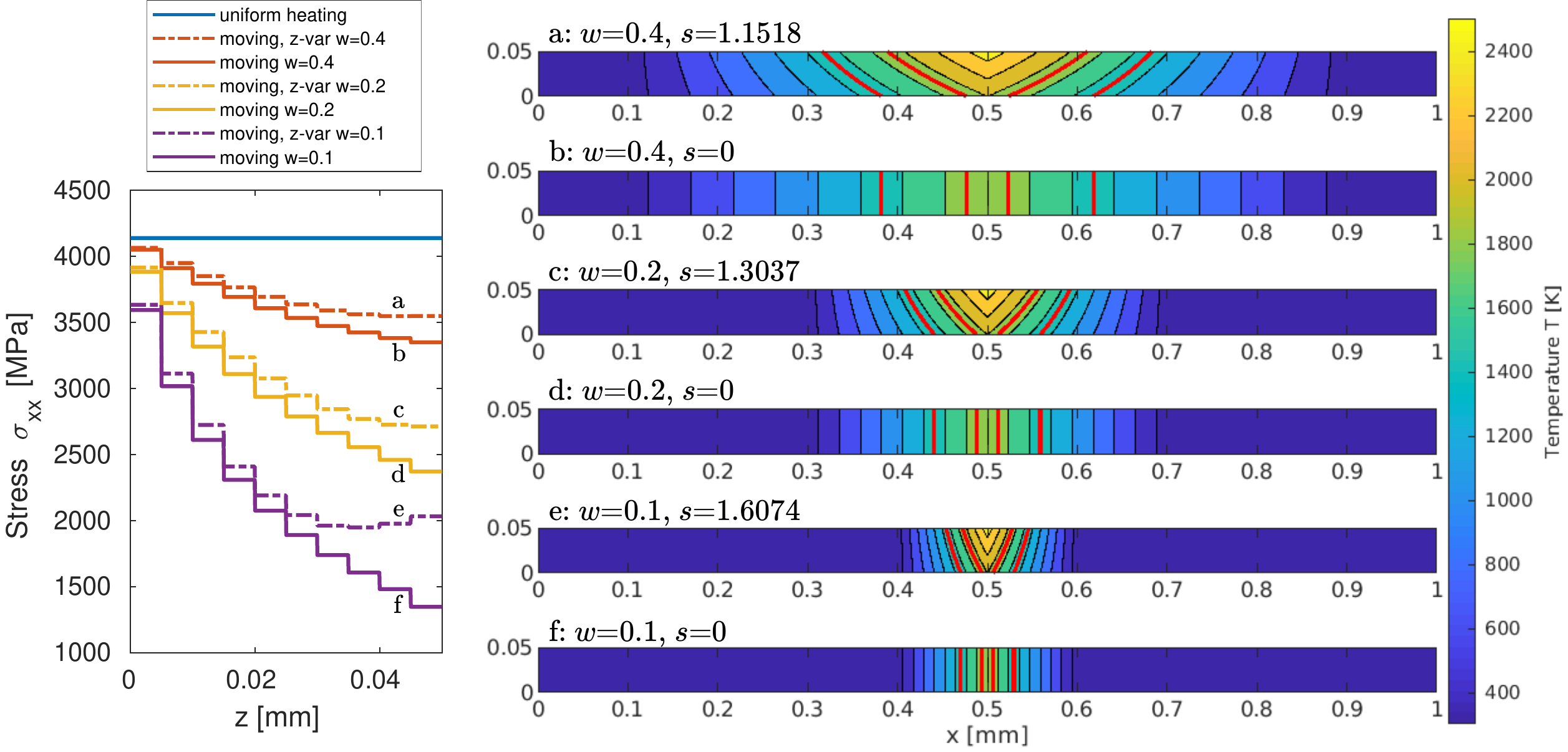}
    \caption{Two-dimensional example. Left: final stress $\sigma_{xx}$ over height $z$ (zero is constrained bottom) at $x=\num{0.5}\,\si{\milli\metre}$ depending on width of temperature peak and (possibly) $z$-variation. Solid lines represent moving temperature profiles, dashed lines represent moving and $z$-varying temperature profiles.  Right: snapshots of temperature profiles $a$--$f$ which induce the associated final stress response on the left. Solidus and liquidus isotherms indicated in red.}
    \label{fig:2d_combined}
\end{figure}

Next, the temperature profile shall vary in $z$-direction as well. Given a time and $x$-dependent function $T_{z0}(x,t)$ for the temperature profile at $z=0$ the following function
\begin{align}
    \label{eq:2d_z_variation}
    T(x,z,t) = \frac{z(s-1)+h}{h}T_{z0}(x,t),
\end{align}
allows to control the variation of the temperature over the height $h=0.05 \,\si{\milli\metre}$. The scaling factor $s$ describes the ratio of the temperature at the upper edge $z=h$ compared to the constrained bottom $z=0$. For better comparison, it is calculated from the width $w$ in \eqref{eq:1d_temperature_peak_profile} as
\begin{align}
    \label{eq:2d_z_variation_scaling}
    s = 1+ \frac{T_\text{max}-T_0}{T_m- T_0}\frac{h}{w},
\end{align}
such that the partial derivative of \eqref{eq:2d_z_variation} with respect to $z$ is comparable to the slope of the moving temperature peak of width $w$ in \eqref{eq:1d_temperature_peak_profile}.

First, the temperature only varies in $z$-direction but is constant in $x$-direction, which is achieved by \eqref{eq:2d_z_variation} and the purely time-dependent function
\begin{align}
    T_{z0}(t) = T_\text{max} \times \begin{cases}
        2t,& \text{if }t \leq 0.5\, \si{\second}\\
        2-2t,& \text{if }t > 0.5\, \si{\second}\\
    \end{cases}.
\end{align}
The scaling factor $s$ is computed from $w$ according to \eqref{eq:2d_z_variation_scaling}.
The results in Figure~\ref{fig:2d_zcut} (right) indicate that a variation only in $z$-direction does not lead to significantly different stresses (compared to the uniform temperature field), even for extreme values such as $w=0.01 \,\si{\milli\metre}$, i.e., a scaling factor of $s>7$ between upper and lower edge temperatures. All material points with the same $z$-coordinate solidify at the same time.

The situation becomes more interesting when the moving peak is combined with a variation in $z$-direction, which is closest to the real temperature profiles expected in PBFAM processes. This is easily achieved by inserting \eqref{eq:1d_temperature_peak_profile} as $T_{z0}$ into \eqref{eq:2d_z_variation} to obtain another kind of temperature profile, see Figure~\ref{fig:2d_combined} (right). The width (in $x$-direction) of the peak again varies between $w \in \lbrace 0.4,0.2,0.1\rbrace\, \si{\kelvin}$. Figure~\ref{fig:2d_combined} (left) shows how the stress $\sigma_{xx}$ varies over the height $z$. Especially for small widths, the additional variation in $z$-direction has an influence on the resulting stress compared to the corresponding cases without $z$-variation. 

We do not claim that the results in this section are directly transferable to a PBFAM process simulation. Still, they suggest that the actual temperature profile might play a more important role than often assumed. In the authors' opinion, the significant differences in residual stress from a uniform and a moving temperature field raise the question how well-suited common simplifications in PBFAM simulation, such as layer-agglomeration, inherent-strain or flashing complete tracks at once, are for an accurate prediction of residual stress distributions.

\section{Multilayer {mesh tying} strategy}
\label{sec:meshtying}
\newcommand{\discret}[1]{\vec{#1}}

\begin{figure}[bt]
    \centering
    \includegraphics[width=.4\linewidth]{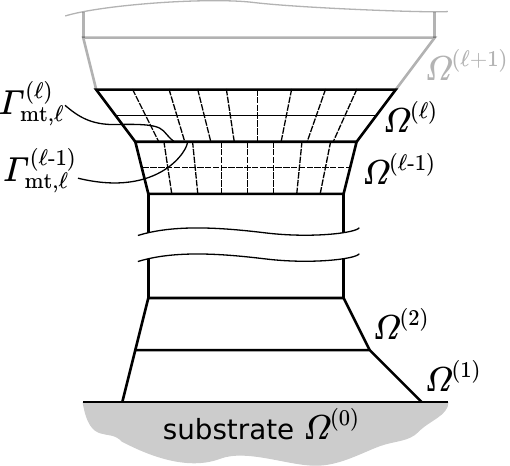}
    \caption{{Mesh tying} concept for multilayer PBFAM simulations.}
    \label{fig:meshtying_concept}
\end{figure}

From the PBFAM process perspective, it is natural to slice the complex part geometries into layers, which may either represent a single powder layer or an upscaled process layer, i.e., an accumulation of several powder layers. It seems that many existing approaches, even in commercial codes, suffer from the need to generate matching meshes between layers for real application scenarios and complex geometries. In this work, we want to propose a flexible, layerwise spatial discretization strategy without the need of matching meshes across the interface, thus allowing for easy mesh generation without distorted elements or large size differences between elements.
Dual mortar methods allow to couple such non-matching FEM meshes in a manner that results in optimal discretization errors (measured in the $L^2$-norm) and that leads to an efficient condensation of the constraint equations from the global system of equations.

This section summarizes the most important aspects of the approach for the application to PBFAM, while a general overview on mortar methods for contact and mesh tying problems is found in \cite{Popp2010, Puso2003}. Figure~\ref{fig:meshtying_concept} shows an exemplary part geometry which is sliced into layers $\Omega^{(\ell)}$ with thickness equal to the powder thickness $h_p$. Each layer can be meshed separately and the finite element nodes may be non-matching at the interface between two layers. The continuity of the primary solution variables across the interface $\Gamma_{\text{mt},\ell}$ is enforced in a weak sense by additional (mesh tying) constraint equations. The $\ell$-th mesh tying interface $\Gamma_{\text{mt},\ell}$ consists of the boundaries $\Gamma_{\text{mt},\ell}^{(\ell-1)}$ (top surface of domain $\Omega^{(\ell-1)}$) and  $\Gamma_{\text{mt},\ell}^{(\ell)}$ (bottom surface of domain $\Omega^{(\ell)}$). In this framework, applying a new powder layer boils down to activating a new mesh tying constraint.

In order to have a continuous part, degrees of freedom (DOFs) associated with nodes on the bottom surface of powder domain  $\Omega^{(\ell)}$ and on the top surface of powder domain $\Omega^{(\ell-1)}$ need to be constrained.
Displacements and temperatures are the primary solution variables of the thermomechanical problem~\eqref{eq:heat_equation} and \eqref{eq:balance_linear_momentum} and thus the necessary constraint equations read,
\begin{align}
\label{eq:mm_layers_u_meshtying}
\vec{u}^{(\ell)} &= \vec{u}^{(\ell-1)},& &\text{on}\ \Gamma_{\text{mt},\ell},\\
\label{eq:mm_layers_T_meshtying}
T^{(\ell)} &= T^{(\ell-1)},& &\text{on}\ \Gamma_{\text{mt},\ell}.
\end{align}
These conditions are enforced by introducing a Lagrange multiplier potential for each constraint:
\begin{align}
    \Pi_{\vec{u},\text{mt}} &= \int_\Gamma \vec{\lambda}_{\vec{u}} \cdot \left(\vec{u}^{(\ell)}-\vec{u}^{(\ell-1)}\right)\,\dd \Gamma_{\text{mt},\ell},\\  \Pi_{T,\text{mt}} &= \int_\Gamma \lambda_T \cdot \left(T^{(\ell)}-T^{(\ell-1)}\right)\,\dd \Gamma_{\text{mt},\ell}.
\end{align}
The total variation of the two potentials is found as
\begin{align}
    \label{eq:variation_lagr_mult_pot_u}
    \delta \Pi_{\vec{u},\text{mt}} &= \int_\Gamma \delta \vec{\lambda}_{\vec{u}}  \cdot \left(\vec{u}^{(\ell)}-\vec{u}^{(\ell-1)}\right)\,\dd \Gamma_{\text{mt},\ell} + \int_\Gamma \vec{\lambda}_{\vec{u}}  \cdot \left(\delta \vec{u}^{(\ell)}-\delta \vec{u}^{(\ell-1)}\right) \,\dd \Gamma_{\text{mt},\ell}\\
    \label{eq:variation_lagr_mult_pot_T}
    \delta \Pi_{T,\text{mt}} &= \int_\Gamma \delta \lambda_T  \cdot \left(T^{(\ell)}-T^{(\ell-1))}\right) \,\dd \Gamma_{\text{mt},\ell} + \int_\Gamma \lambda_T \cdot  \left(\delta T^{(\ell)}-\delta T^{(\ell-1)}\right)\,\dd \Gamma_{\text{mt},\ell}
\end{align}
The space and time dependency of $\vec{u}$, $T$, their variations $\delta \vec{u}$ and $\delta T$, and the Lagrange multipliers $\vec{\lambda}_{\vec{u}}$ and $\lambda_T$ is not written out in equations \eqref{eq:mm_layers_u_meshtying}--\eqref{eq:variation_lagr_mult_pot_T} for brevity. The Lagrange multipliers $\lambda_\square$ and their variations $\delta \lambda_\square$ are chosen from $\mathcal{M}_\square$, the dual space of the trace space of $\mathcal{V}_\square$, i.e., $\mathcal{M}_\square = \mathcal{H}^{-1/2}(\Gamma_\text{mt})$, where $\square \in \lbrace \vec{u}, T\rbrace$.
The first term in \eqref{eq:variation_lagr_mult_pot_u} and \eqref{eq:variation_lagr_mult_pot_T} represents the original mesh tying constraint and the second term contributes to the weak forms \eqref{eq:heat_equation_weak} and \eqref{eq:balance_linear_momentum_weak}. From dimensional analysis it becomes apparent that the unknown structural Lagrange multiplier can be interpreted as a force vector acting on the interface, $\vec{\lambda}_{\vec{u}} = \vec{t}_\text{mt}^{(\ell)} = -\vec{t}_\text{mt}^{(\ell-1)}$, and the thermal Lagrange multiplier as a heat flux across the interface, $\lambda_T = q_\text{mt}^{(\ell)} = - q_\text{mt}^{(\ell-1)}$. The Lagrange multiplier fields are discretized with special dual shape functions that allow for an efficient condensation of the Lagrange multiplier DOFs from the global system of equations~\cite{Puso2003}. After spatial discretization of all primary variable fields, the consistent link between discrete nodal DOFs on both sides of the interface is established via constant mortar matrices $\discret{D}_\square$ and $\discret{M}_\square$~\cite{Popp2010}, which can be combined into constant projector matrices $\discret{P}_\square=\discret{D}_\square^{-1}\discret{M}_\square$. The projector matrices map discrete solution increments (from the nonlinear solution procedure),
\begin{align}
\label{eq:mm_projector_d_inc}
\Delta\discret{d}^{(\ell)}& = \discret{P}_{\vec{u}}\Delta\discret{d}^{(\ell-1)}, \quad \text{on}\ \Gamma_{\text{mt},\ell}\\
\label{eq:mm_projector_T_inc}
\Delta\discret{T}^{(\ell)} &= \discret{P}_{T}\Delta\discret{T}^{(\ell-1)},  \quad \text{on}\ \Gamma_{\text{mt},\ell},
\end{align}
where $\Delta \discret{d}$ and $\Delta\discret{T}$ are increments of the discrete solution vectors of discrete displacements $\discret{d}$ and discrete temperatures $\discret{T}$.
Given that the mesh tying constraints \eqref{eq:mm_layers_u_meshtying} and \eqref{eq:mm_layers_T_meshtying} hold initially, these incremental mappings will enforce the constraints also in later configurations. Typically this is easy to ensure in problem settings which utilize a tied interface from the beginning. However, in PBFAM applications, new powder layers are added over time, thus introducing new mesh tying constraints. In general, a new powder layer's initial temperature and displacement will not conform to the, already processed, last layer.
A simple approach to initialize temperature and displacements on side $\Gamma_{\text{mt},\ell}^{(\ell)}$ of a newly added layer is to apply mappings \eqref{eq:mm_projector_d_inc} and \eqref{eq:mm_projector_T_inc} to the solution vectors on $\Gamma_{\text{mt},\ell}^{(\ell-1)}$, viz.
\begin{align}
    \label{eq:mortar_init_mapping}
\discret{d}^{(\ell)} \leftarrow \discret{P}_{\vec{u}}\discret{d}^{(\ell-1)},\quad
\discret{T}^{(\ell)} \leftarrow \discret{P}_{T}\discret{T}^{(\ell-1)}\quad \text{on }  \Gamma_{\text{mt},\ell}.
\end{align}
In essence, the nodal solution at the bottom surface of a newly added layer is set to the (consistently mapped) nodal solution at the top surface of the last processed layer. After this initialization the meshes are correctly tied and the primary solution variables are continuous across the interface. 
The remaining  DOFs in the newly added domain $\Omega^{(\ell)} \backslash \Gamma_{\text{mt},\ell}^{(\ell)}$ are set as follows: temperature DOFs are set to the initial temperature $T_0$, while displacement DOFs are initialized to zero. The first time step after a new mesh tying interface is activated will solve the (static) mechanical equilibrium equation \eqref{eq:balance_linear_momentum} for a consistent displacement state.
\begin{remark}
Technically, the direct mapping \eqref{eq:mortar_init_mapping} violates conservation of energy when adding a new powder layer: temperature and displacement on $\Gamma_{\text{mt},\ell}^{(\ell)}$ are modified without an associated external force. An alternative initialization of the temperature field for element-birth methods is discussed in~\cite{Michaleris2014} and could be transferred to the proposed mesh tying approach. However, if the cooldown phase after processing of one layer is long enough for the top surface to reach the initial temperature, the energy error from the temperature modification vanishes. The error introduced by modification of the displacements of a newly added layer is generally small due to the low stiffness of the newly added powder, i.e., possible artificial strains in the new layer only cause a small stress response. Furthermore, the material formulation from Section \ref{sec:mechanical_material_law} ensures that these artificial stresses vanish after melting.
\end{remark}

\section{Three-dimensional numerical examples}
\label{sec:three_dim_examples}
This section presents larger three-dimensional numerical examples that are simulated with the research code BACI~\cite{BACI}, jointly developed at the Institute for Computational Mechanics.

\subsection{Single tracks per layer}
\label{sec:3d_single_tracks}

\begin{figure}
    \centering
    \includegraphics[width=.55\linewidth]{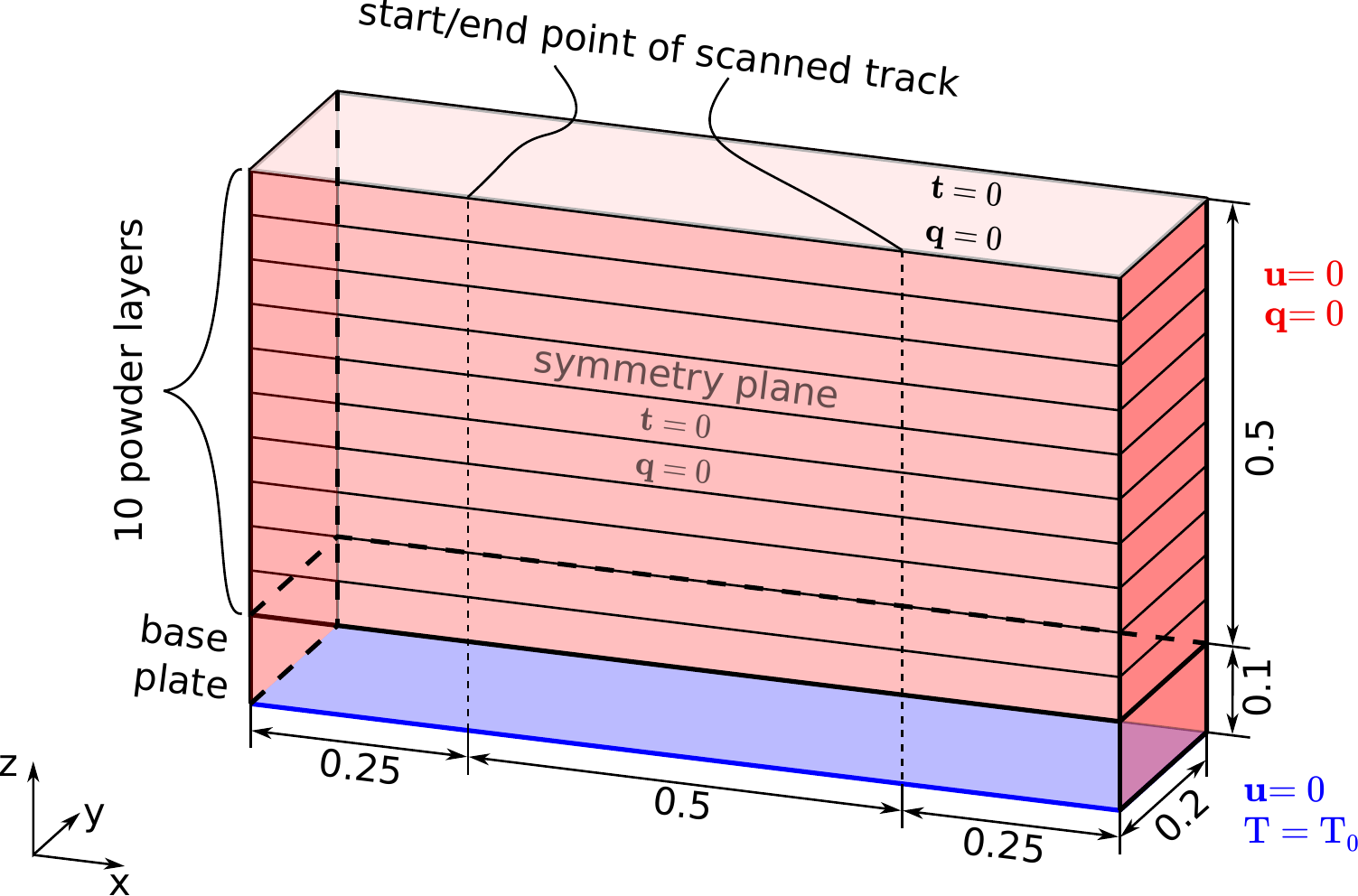}
    \caption{Geometry, boundary conditions and simulation setup of ten layer example. Dimensions in \si{\milli\metre}.}
    \label{fig:10layer_setup}
\end{figure}

The first three-dimensional example investigates ten processed powder layers with a single track per layer each on top of a solid base plate. A schematic of the geometry and the boundary conditions is shown in Figure~\ref{fig:10layer_setup}. The computational domain consists of a \num{0.1} \si{\milli\metre} high base plate that has a prescribed temperature $T_0=\num{303} \,\si{\kelvin}$ and zero displacements $\vec{u}=\vec{0}$ at the bottom (indicated in blue in Figure~\ref{fig:10layer_setup}). This temperature $T_0$ is also used as the initial temperature throughout the domain and in all newly activated powder layers. The powder layers of height $h_p$ are connected via the mesh tying algorithm introduced in the last section as soon as they are activated. The tracks are scanned atop each other with a unidirectional or serpentine pattern, i.e., the laser tracks are either oriented in the same direction in all layers (unidirectional) or the orientation alternates between successive layers (serpentine). To save computational resources, a symmetry plane along the laser track center line is used to half the size of the computational domain.  All lateral faces (indicated in red) are assumed to be thermally insulating ($\vec{q}=0$) and subject to homogeneous Dirichlet boundary conditions.. The built part will not attach to these boundaries for the chosen laser beam path and diameter and, since the remaining powder close to the boundary is modeled with low stiffness and conductivity, the zero displacement boundary conditions will not influence the results in the final part geometry significantly.

\begin{table}
    \centering
\caption{Material and simulation parameters for three-dimensional examples.}
\label{tab:params_3d}
\scriptsize
\begin{tabular}{llll}
    \toprule
    Quantity & Description & Value & Unit \\
    \midrule
    $E_s$ & Young's modulus in solid & \num{200} & \si{GPa} \\
    $E_p,E_m$ & (Artificial) Young's modulus in powder and melt & \num{2} & \si{GPa} \\
    $\nu$ & Poisson ratio & \num{0.3} & -- \\
    $\alpha_T$ & Coefficient of thermal expansion & \num{15e-6} & \si{\per\kelvin}\\
    \midrule
    $c_s$ & Volumetric specific heat, solid & \num{4.25} & \si{\mega\joule\per\cubic\metre\per\kelvin}\\
    $c_p$ & Volumetric specific heat, powder & \num{2.98} & \si{\mega\joule\per\cubic\metre\per\kelvin}\\
    $c_m$ & Volumetric specific heat, melt & \num{5.95} & \si{\mega\joule\per\cubic\metre\per\kelvin}\\
    $k_s,k_m$ & Conductivity in solid and melt & \num{20} & \si{\watt\per\metre\per\kelvin}\\
    $k_p$ & Conductivity in powder & \num{0.2}@\num{200}, \num{0.3}@\num{1600}  & \si{\watt\per\metre\per\kelvin}@\si{\kelvin}\\
    $h_m$ & Latent heat of fusion & \num{2.18} & \si{\giga\joule\per\cubic\metre}\\
    $T_0$ & Initial/reference temperature & \num{303} & \si{\kelvin}\\
    $T_s$ &Solidus temperature & \num{1500} & \si{\kelvin} \\
    $T_l$ &Liquidus temperature & \num{1900} & \si{\kelvin}\\

    \midrule
    $\rho_h$ & Hemispherical reflectivity & \num{0.7} & --\\
    $\beta_h$ & Extinction coefficient & \num{60} & \si{\per\milli\metre}\\
    $h_p$ & Powder layer thickness & \num{50} & \si{\micro\metre}\\
    $W_e$ & Effective laser power & \num{30} & \si{\watt}\\
    $R$ & Effective laser radius & \num{0.08} & \si{\milli\metre}\\
    $v_\text{scan}$ & Laser scan speed & \num{100} & \si{\milli\metre\per\sec}\\
    \bottomrule
\end{tabular}
\end{table}

Every scanned track (pure scanning time of 0.005 \si{\second}) is followed by a cooldown time of $\num{1.0}\,\si{\second}$ (which is long enough to reach a homogeneous temperature state close to the initial value $T_0$ in good approximation) during which significant residual stresses form.
After all ten layers have been scanned, the process of cutting the part from the base plate can be simulated as follows: The boundary conditions on the red surfaces are removed everywhere except on the leftmost surfaces (with normal vector $(-1,0,0)^T$) to avoid rigid body modes. The mesh tying constraint between the base plate and the first layer is removed. A structural analysis is run to obtain a static equilibrium solution for the stress and deformation state after detaching the part from the base plate.

Material parameters for the thermal and mechanical problem as well as other processing parameters are listed in Table~\ref{tab:params_3d}.
A time step of $\Delta t = \num{5e-6}\,\si{\second}$ is used for the scanning phase and the initial cooling phase of $0.002\,\si{\second}$, while the remaining $0.998\,\si{\second}$ of cooling use a time step of $\Delta t = \num{1e-3}\,\si{\second}$. The whole domain is discretized with \num{61440} linear, hexahedral 8-node finite elements with edge size $h^\text{ele}=\frac{h_p}{4} = \num{12.5}\,\si{\micro\metre}$.

\begin{figure}[bt]
    \centering\includegraphics[width=.95\linewidth]{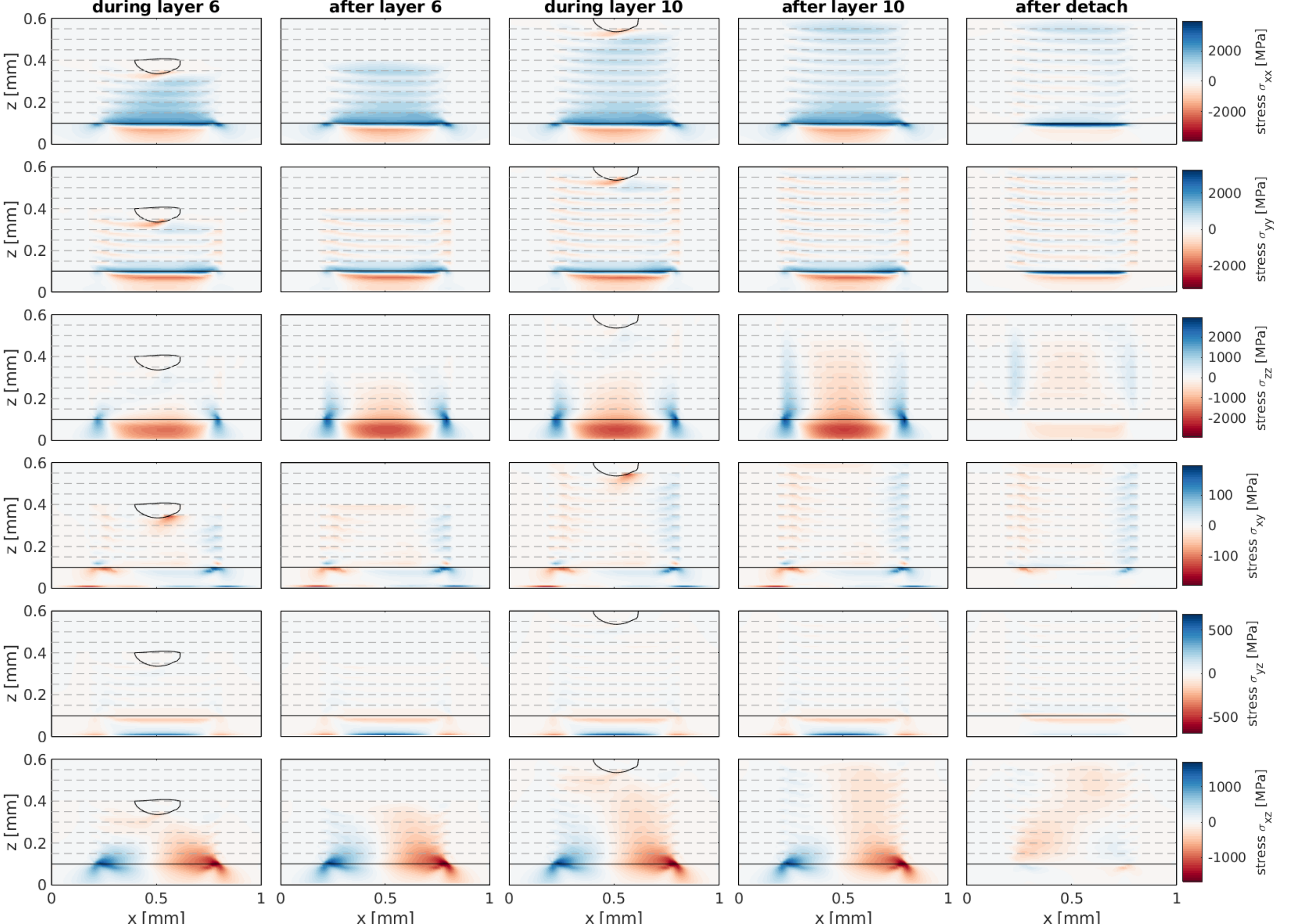}
    \caption{Unidirectional scanning pattern: distribution of different stress components in symmetry plane at selected time steps. Blue color represents tension, red color represents compression. Liquidus isotherm (solid black line) approximates melt pool shape.}
    \label{fig:10layer_stress_xz_uniform}
\end{figure}

\begin{figure}[bt]
    \centering\includegraphics[width=.95\linewidth]{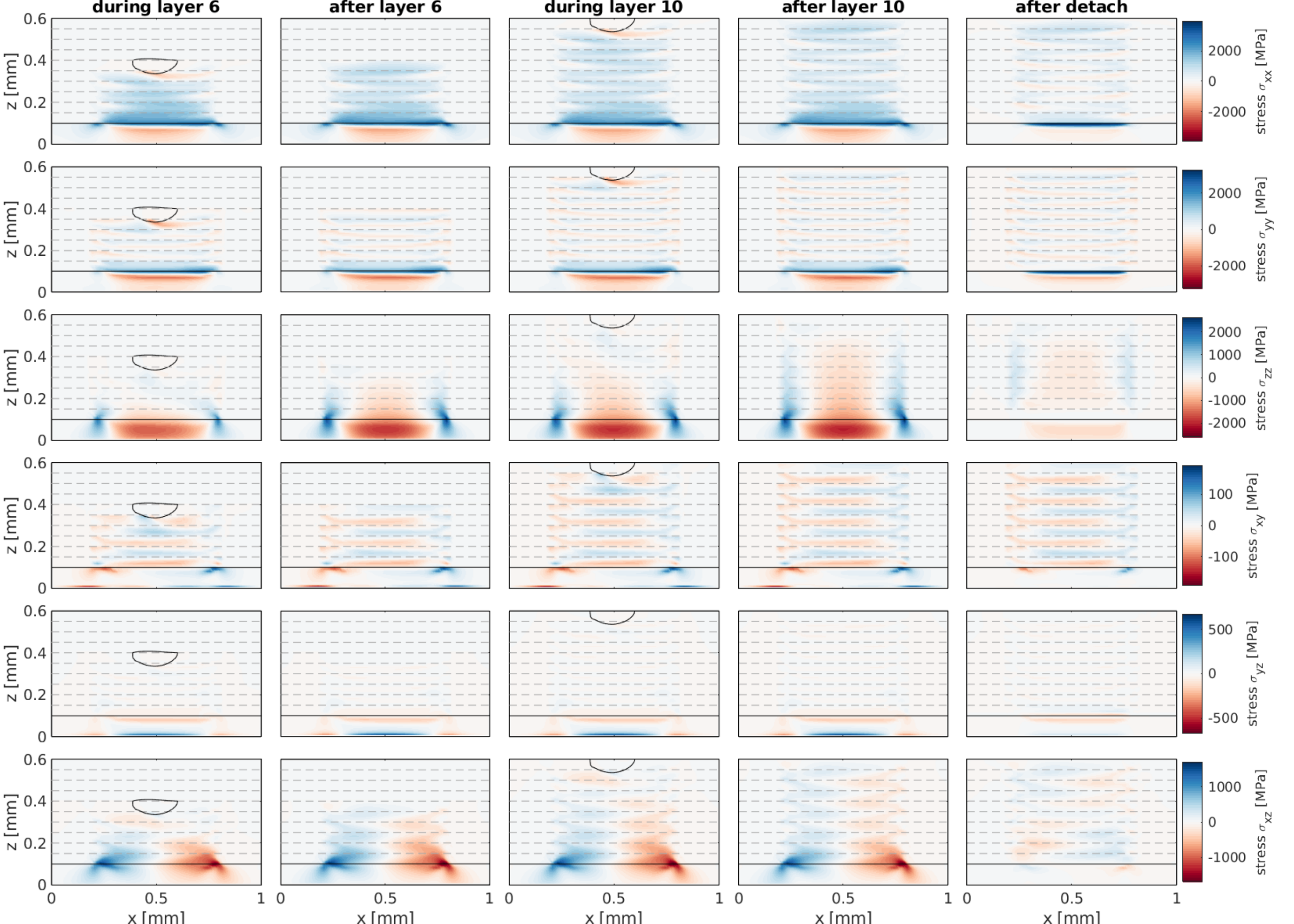}
    \caption{Serpentine scanning pattern: distribution of different stress components in symmetry plane at selected time steps. Blue color represents tension, red color represents compression. Liquidus isotherm (solid black line) approximates melt pool shape.}
    \label{fig:10layer_stress_xz_serpentine}
\end{figure}
Figure~\ref{fig:10layer_stress_xz_uniform} shows an overview of the six components of the stress tensor at selected points in time: the first and third column show stresses \textit{during} scanning of layer 6 and 10 with a comparable relative laser position. The second and fourth column show stresses \textit{after} these layers have been cooled down. The fifth column shows the residual stress after the part is detached from the base plate. The laser paths of the individual tracks remain clearly visible in the plots of the $\sigma_{xx}$ component (first row).
In agreement with the literature this stress in scanning direction is a large tensile stress \cite{Bruna-Rosso2020, Carraturo2021}. The highest stresses can be observed in the heavily constrained base plate. Specifically, it can be observed how geometrical compatibility during cool down lead to high tensile normal stresses $\sigma_{xx}$ in the first track (and remolten portion of the base plate) and high compressive normal stresses $\sigma_{xx}$ in the (non-molten) lower part of the build plate. In the last row of Figure \ref{fig:10layer_stress_xz_uniform} the shear stresses $\sigma_{xz}$ can be observed, which are - according to the mechanical equilibrium equation \eqref{eq:balance_linear_momentum} - necessary to balance gradients of the normal stress $\sigma_{xx}$ at the beginning and end of a track, especially close to the base plate. These $\sigma_{xz}$ shear stresses, in turn, also induce $\sigma_{zz}$ normal stresses. Note, that the different stress components at many locations exceed the yield stress of typical materials used in PBFAM, hence the inclusion of an elasto-plastic material model, as already sketched in Section~\ref{sec:material_mathematical}, is desirable for future work. The inclusion of such material models, or even of more elaborate microstructure-informed constitutive laws, will allow for a detailed quantitative validation based on experimental results.

\begin{figure}
    \centering
    \includegraphics[width=.95\linewidth]{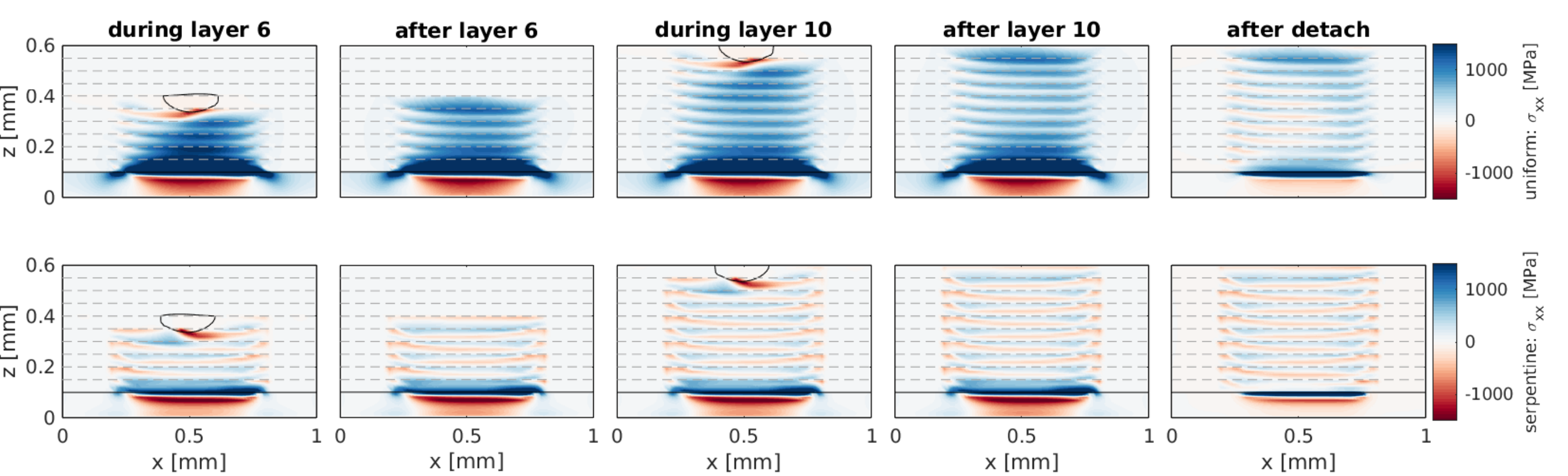}
    \caption{Stress distribution of $\sigma_{xx}$ in symmetry plane at selected time steps for unidirectional (first row) and serpentine (second row) scanning pattern. Blue color represents tension, red color represents compression. Liquidus isotherm (solid black line) approximates melt pool shape. For improved visibility of the relevant stresses the color bar is rescaled.}
    \label{fig:10layer_stress_xz_uniform_serpentine_rescaled}
\end{figure}

All stresses decrease in their magnitude when the part is cut from the base plate. Figure~\ref{fig:10layer_stress_xz_serpentine} shows the resulting stress distribution for an alternating, serpentine scanning pattern. The absolute values are comparable to Figure~\ref{fig:10layer_stress_xz_uniform}. To better highlight the influence of the different scanning patterns, Figure~\ref{fig:10layer_stress_xz_uniform_serpentine_rescaled} visualizes the normal stresses $\sigma_{xx}$ for the unidirectional and serpentine scanning pattern with a rescaled color map.

\begin{figure}
    \centering
    \includegraphics[width=.95\linewidth]{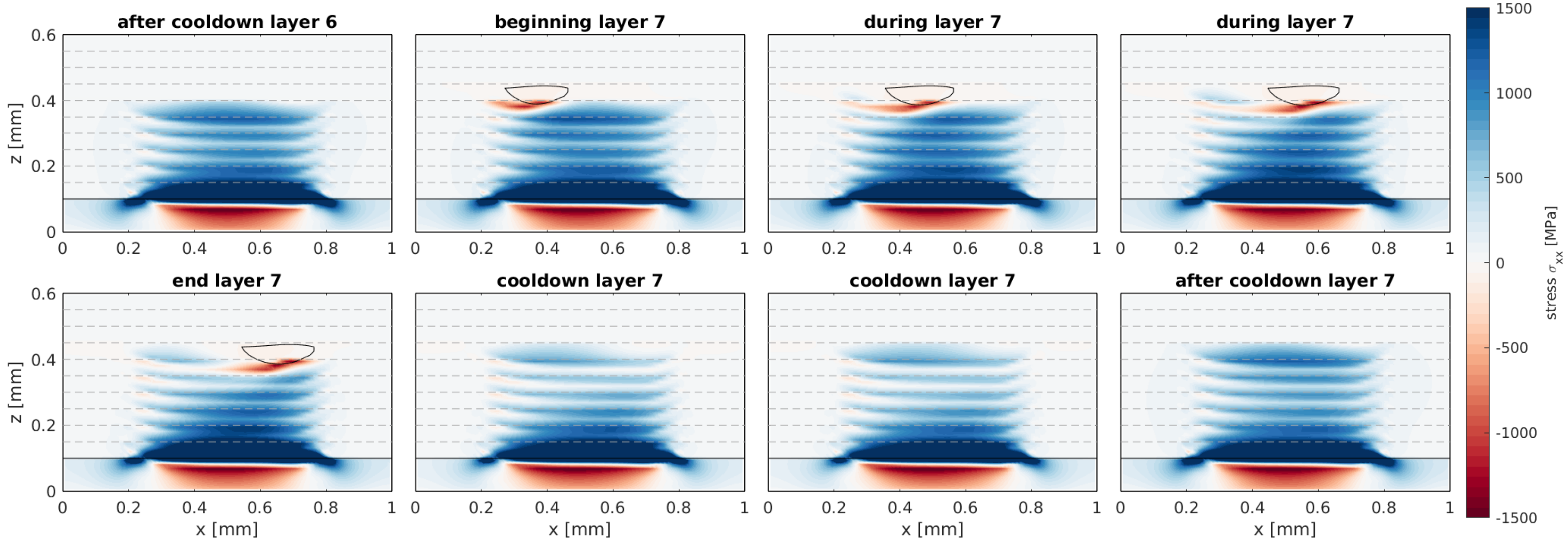}
    \caption{Unidirectional scanning pattern: detailed view of the evolution of the longitudinal stress $\sigma_{xx}$ in the symmetry plane during scanning of layer 7. Blue color represents tension, red color represents compression. Liquidus isotherm (solid black line) approximates melt pool shape. Time progresses from left to right and top to bottom. For improved visibility of the relevant stresses the color bar is rescaled.}
    \label{fig:10layer_stressx_detail_layer7}
\end{figure}

Selected time steps are plotted in Figure \ref{fig:10layer_stressx_detail_layer7} for a more detailed view of the stress evolution of $\sigma_{xx}$ during scanning of track 7 with a unidirectional pattern. The following considerations start from a fully cooled layer 6, which exhibits a tensile stress throughout the consolidated material.
The subsequent snapshots (from left to right) in the first and second row of Figure \ref{fig:10layer_stressx_detail_layer7} show the processing of track 7. To visualize the melt pool size, the temperature isoline corresponding to the liquidus temperature $T_l$ is depicted in these snapshots (solid black line). While no stresses occur in the powder material of layer 7 in front of the laser, the thermally induced volume expansion during heating leads to negative (compressive; red color) stresses in the solid material of the previously processed track 6, mostly pronounced in the direct vicinity of the $T_l$-isoline. As desired, these stresses rapidly drop to zero in the narrow temperature interval $T \in \lbrack T_s ; T_l \rbrack$ such that no visible stresses remain in the melt pool domain.
This strong gradient between vanishing stresses in the melt pool and high compressive stresses in the solid material beneath remains after solidification and is superimposed by additional tensile stress contributions due to the thermally induced volume shrinkage during cooling. After cooling  down (see snapshot at bottom right of Figure \ref{fig:10layer_stressx_detail_layer7}), this process results in high tensile stresses in the upper, re-molten part of track 6, and stresses close to zero in its lower part. The same characteristics are observed in the previously processed tracks below. Even though the base plate has the same stiffness as the solidified tracks above, this characteristic band structure of the normal stresses is much more pronounced for the first track, i.e., the highest overall tensile stresses occur in the first track, accompanied by compressive stresses of comparable magnitude in the base plate below. This observation can be explained by the fact that the base plate is initially stress-free, while solidified melt tracks are subject to tensile stresses after cooling down (e.g., snapshot at bottom right of Figure \ref{fig:10layer_stressx_detail_layer7}), which partly compensate the compressive stresses arising from thermal expansion when processing the subsequent track above. The tensile stresses in layer 1 are transferred to the base plate as concentrated tensile stresses in the vicinity of the part edges, which may lead to notch effects in practice. It should be mentioned once more that the displacement Dirichlet boundary conditions on the lateral faces of the base plate are not responsible for the high stresses and results are almost identical if these conditions are replaced with zero Neumann boundary conditions.

\begin{figure}
    \centering
    \includegraphics[width=.95\linewidth]{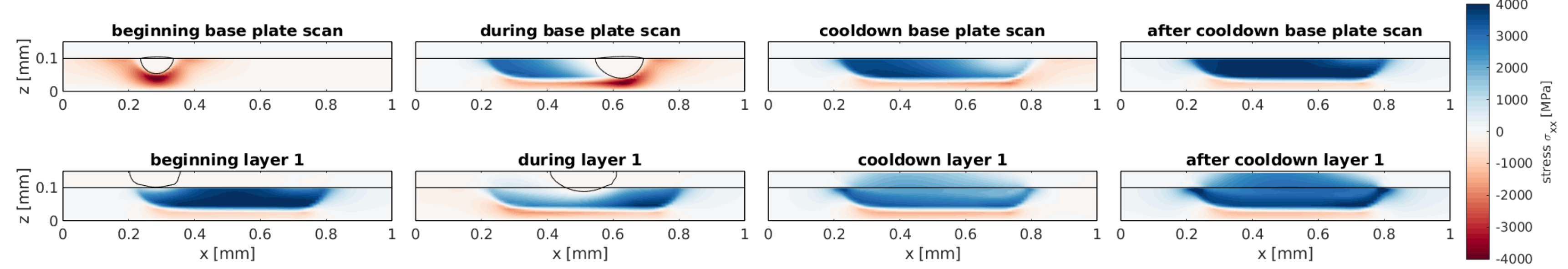}
    \caption{Modified scenario: scanning base plate and layer 1. Detailed view of the evolution of the longitudinal stress $\sigma_{xx}$ in the symmetry plane. Blue color represents tension, red color represents compression. Liquidus isotherm (solid black line) approximates melt pool shape. Time progresses from left to right and top to bottom. For improved visibility of the relevant stresses the color bar is rescaled.}
    \label{fig:baseplate_scan}
\end{figure}

\begin{figure}
    \centering
    \includegraphics[width=.95\linewidth]{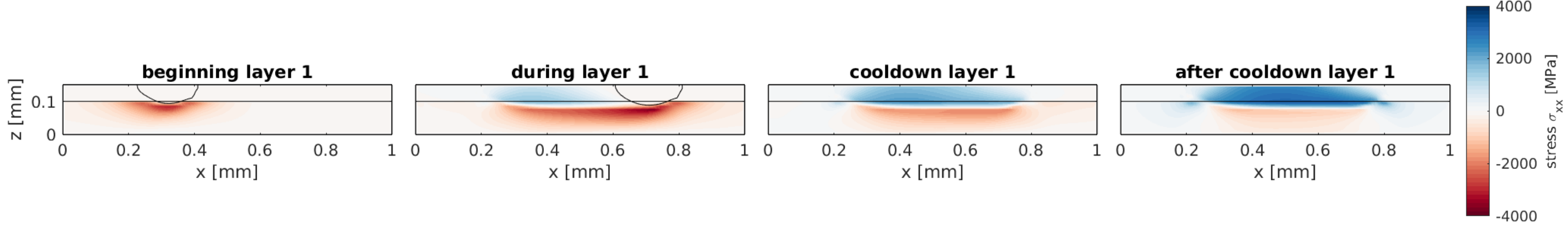}
    \caption{Uniform scanning pattern: detailed view of the evolution of the longitudinal stress $\sigma_{xx}$ in the symmetry plane during scanning of layer 1. Blue color represents tension, red color represents compression. Liquidus isotherm (solid black line) approximates melt pool shape. Time progresses from left to right. For improved visibility of the relevant stresses the color bar is rescaled.}
    \label{fig:10layer_layer1_only}
\end{figure}

Finally, we investigate a slightly modified scenario, where the base plate is scanned first with the same heat source as the powder layers. Figure~\ref{fig:baseplate_scan} shows the evolution of stress $\sigma_{xx}$. Large tensile stresses are visible in the scanned regions of the base plate after cooldown (top right in Figure~\ref{fig:baseplate_scan}). The subsequent processing of layer 1, using a unidirectional pattern, leads to remelting in the uppermost region of the base plate, which causes a reduction of the tensile stress after the next cooldown phase. The strong gradient between tensile and compressive stresses, which was observed close to the interface between base plate and layer 1 in Figure~\ref{fig:10layer_stressx_detail_layer7}, shifts down in Figure~\ref{fig:baseplate_scan} while the magnitude stays roughly the same. 
For comparison, Figure~\ref{fig:10layer_layer1_only} shows a detailed view of the evolution of stress $\sigma_{xx}$ when the first layer is scanned atop a stress-free (not scanned) base plate, i.e., the scenario that was initially investigated in this section. The stress distribution in layer 1 after cooldown for this case (right-most plot in Figure~\ref{fig:10layer_layer1_only}) is equal to the stress distribution in layer 1 after cooldown with a pre-processed base plate (bottom right in Figure~\ref{fig:baseplate_scan}), which demonstrates that the chosen pre-processing of the base plate does not influence the stress in layer 1.

\subsection{Multiple tracks per layer}
\label{sec:3d_multiple}

\begin{figure}[bt]
    \centering
    \includegraphics[width=.9\linewidth]{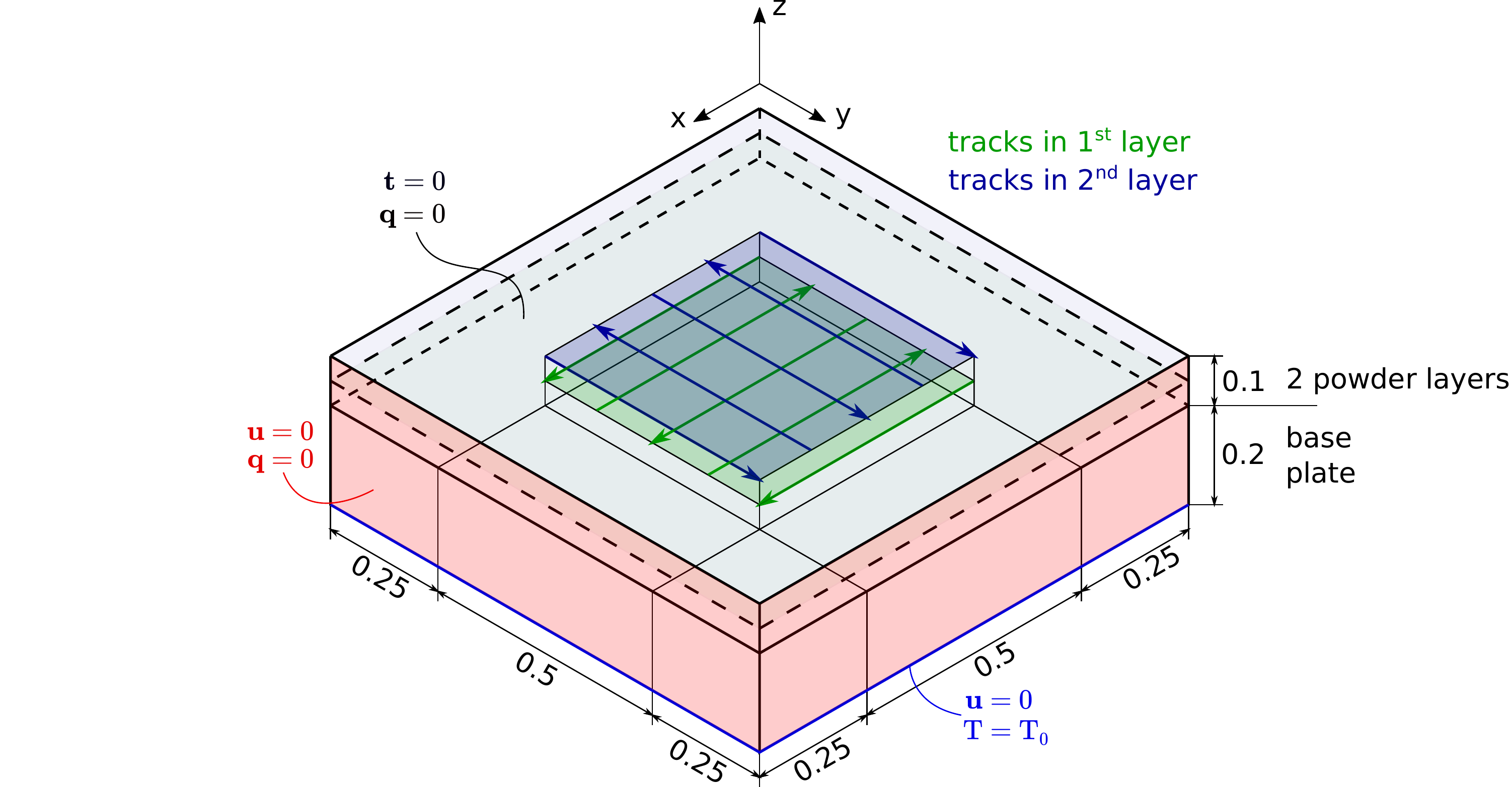}
    \caption{Geometry, boundary conditions and simulation setup for two layer example with multiple tracks per layer (indicated in green for layer 1 and blue for layer 2). Dimensions in \si{\milli\metre}.}
    \label{fig:multitrack_setup}
\end{figure}

\begin{figure}[bt]
    \includegraphics[width=.95\linewidth]{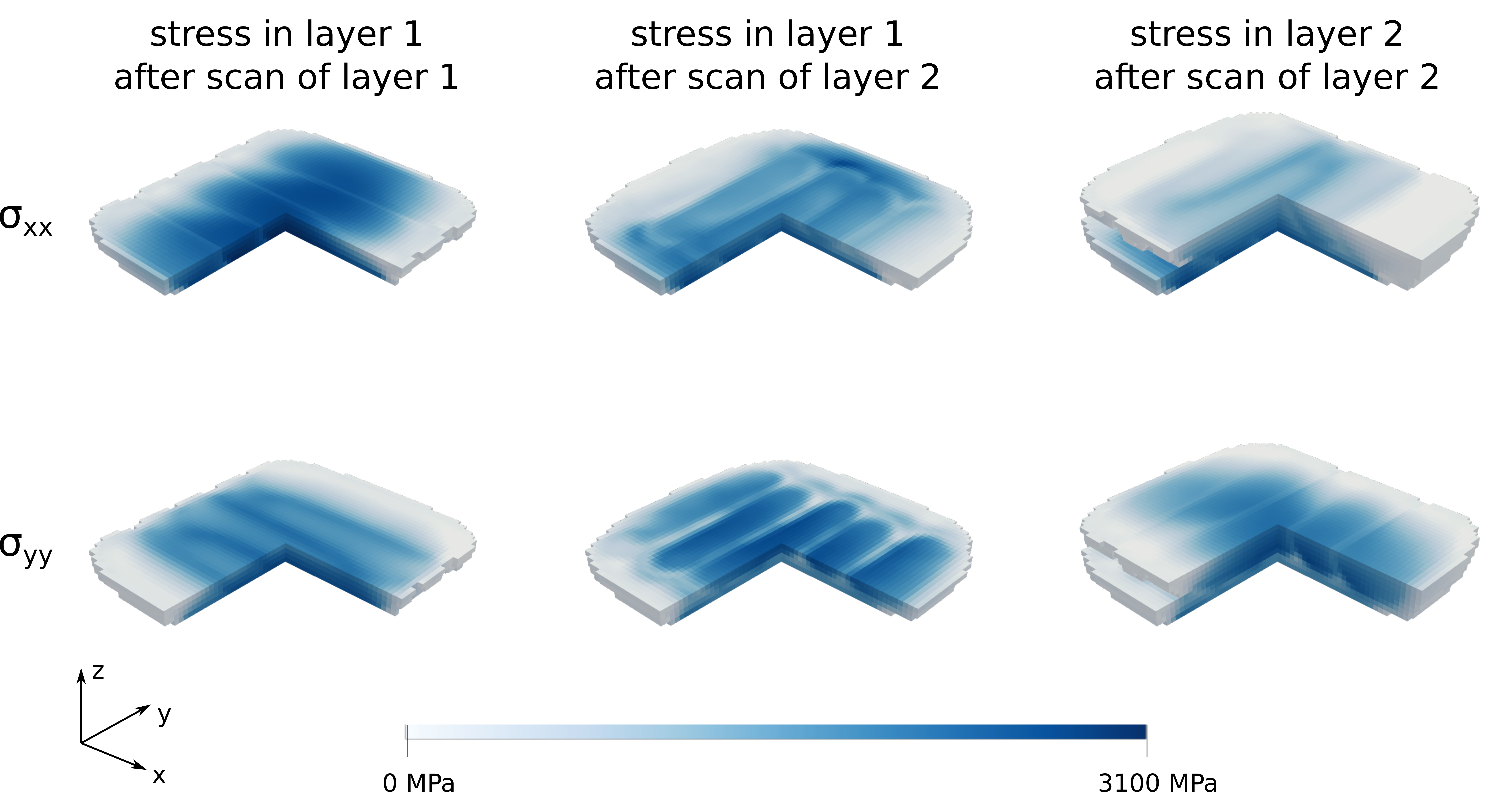}
    \caption{Multiple tracks per layer: stress distribution on consolidated part geometry after scanning layer 1 (first column) and layer 2 (second and third column). A quarter of the geometry is cut out for visualization purposes.}
    \label{fig:multitrack_stress_cutout}
\end{figure}
In the second three-dimensional example, a closer look is taken at the classic in-plane serpentine scanning pattern. The setup is shown in Figure~\ref{fig:multitrack_setup}: two powder layers are deposited on top of a base plate and five tracks following a serpentine pattern are scanned in each layer. Afterwards the domain is allowed to cool down for $\num{1.0}\, \si{\second}$. The boundary conditions are essentially the same as in the last example and the material parameters from Table~\ref{tab:params_3d} are reused. As before, 
a time step of $\Delta t = \num{5e-6}\,\si{\second}$ is used for the scanning phase and the initial cooling phase of $0.002\,\si{\second}$, while the remaining $0.998\,\si{\second}$ of cooling use a time step of $\Delta t = \num{1e-3}\,\si{\second}$. The domain is discretized with \num{153600} linear, hexahedral 8-node finite elements with edge size $h^\text{ele}=\frac{h_p}{4} = \num{12.5}\,\si{\micro\metre}$.

In this example, we look at the stress distribution after each layer was processed. Figure~\ref{fig:multitrack_stress_cutout} visualizes the stresses $\sigma_{xx}$ and $\sigma_{yy}$ in the consolidated material, which represents the built part. One quarter of the consolidated volume is cut out for visualization of the stress distribution in $z$-direction. In the first layer, the laser travels in $x$-direction, which is clearly reflected in the stress distributions shown in the first column of Figure~\ref{fig:multitrack_stress_cutout}.
When the second layer is applied and scanned in $y$-direction, the melt pool penetrates into the first layer. Therefore, after the second layer was scanned, the stress distribution in the first layer reflects the laser beam movement in $y$-direction, see the second column in Figure~\ref{fig:multitrack_stress_cutout}).  At the same time, stresses in the second layer also align with the $y$ direction, see the third column in Figure~\ref{fig:multitrack_stress_cutout}. Overall, only tensile stresses remain after the final cool down.

\begin{figure}[btp]
    \centering
    \includegraphics[width=.7\linewidth]{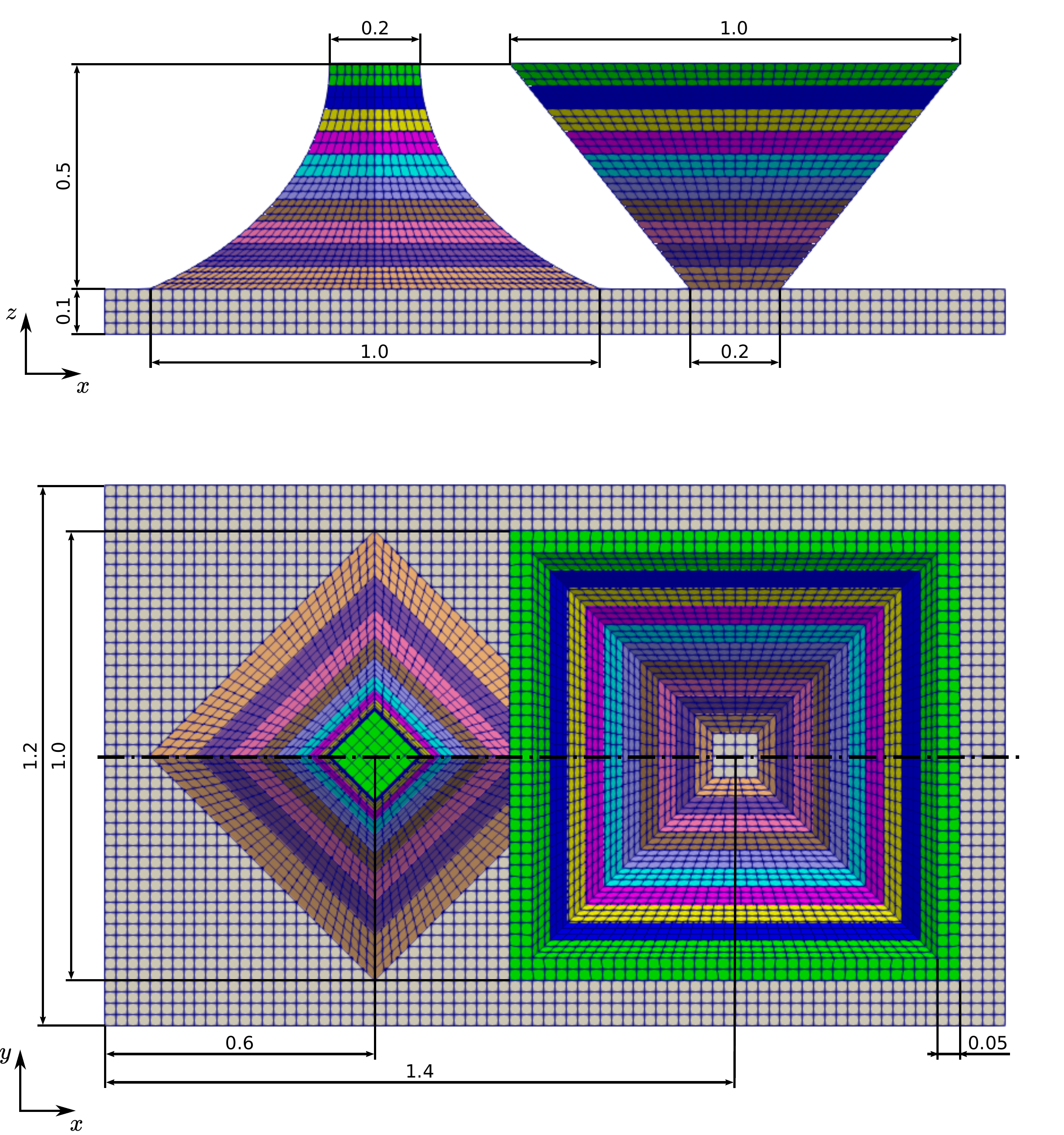}
    \caption{Geometry of pyramids example. Dimensions in \si{\milli\metre}.}
    \label{fig:pyramids_geometry}
\end{figure}

\begin{figure}[btp]
    \centering
    \includegraphics[width=.95\linewidth]{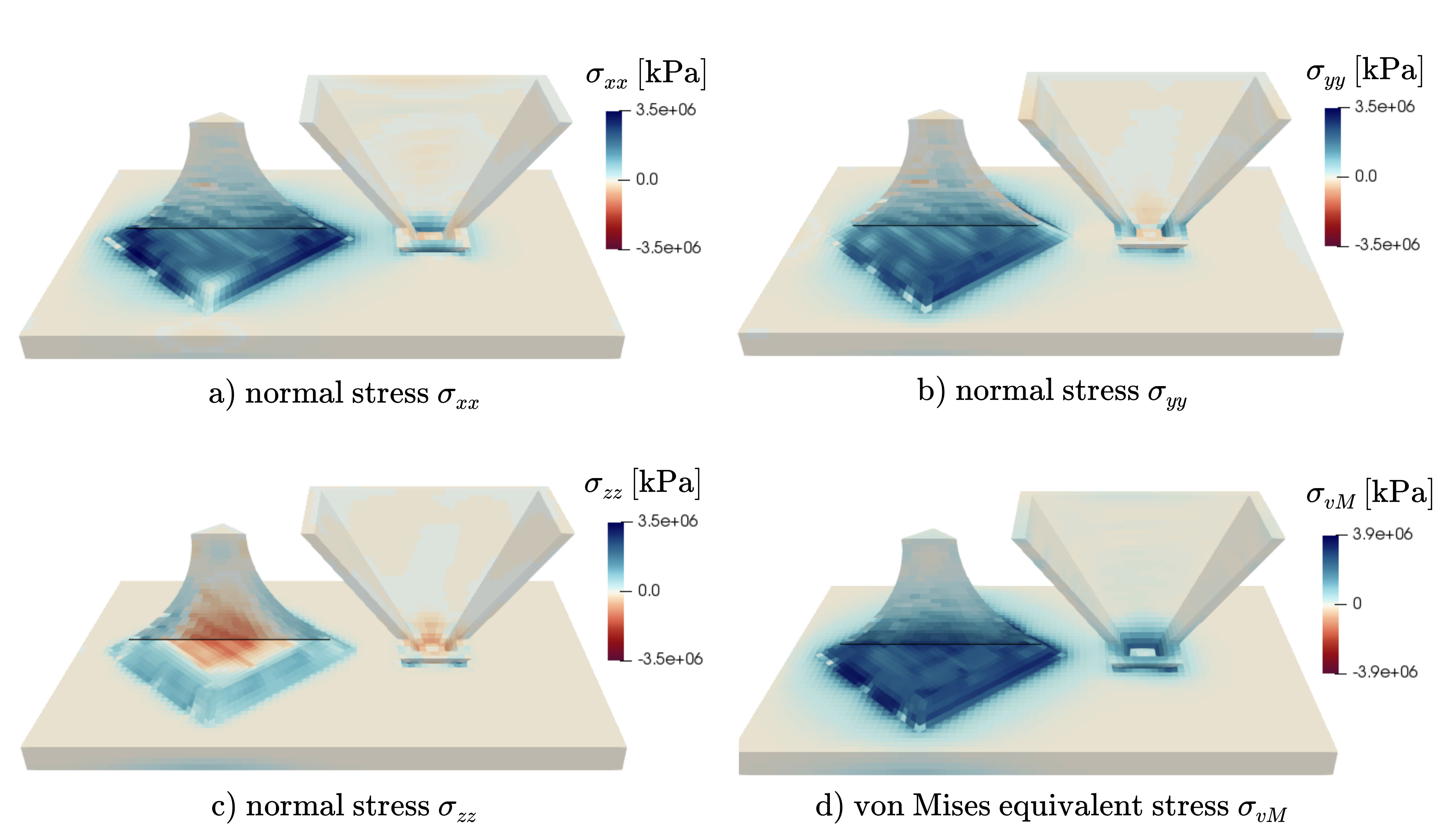}
    \caption{Two pyramids example: stress distribution in pyramids after final cooldown. Normal stresses a) $\sigma_{xx}$, b) $\sigma_{yy}$, c) $\sigma_{zz}$, and d) von Mises equivalent stress $\sigma_{vM}$. Part of the geometry is cut out above first powder layer and at $y=0.6\,\si{\milli\metre}$ for visualization of stress inside the pyramid.}
    \label{fig:double_pyramid_cuts}
\end{figure}

\begin{figure}[btp]
    \centering
    \includegraphics[width=.7\linewidth]{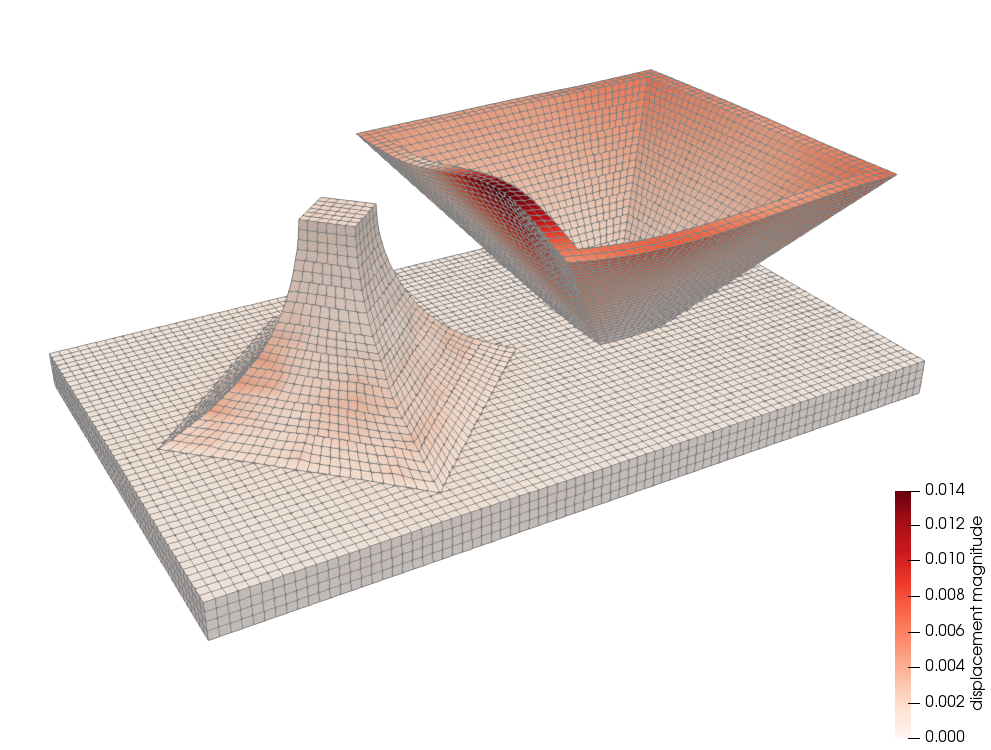}
    \caption{Two pyramids example: displacement magnitude after final cooldown displayed on warped geometry. The distortion is scaled up by a factor of 5.}
    \label{fig:double_pyramid_displacement_warp}
\end{figure}

\subsection{Two pyramids example}
\label{sec:3d_pyramids}
The final numerical example investigates a more complex geometry. Yet, the geometry is defined simple enough so that it can easily be adopted by other researchers. The powder material that is not supposed to contribute to the final part is not included, an assumption which is justified by the low thermal conductivity and stiffness of the powder. The final part's cross section changes drastically over the layers which demonstrates the flexibility of the mortar mesh tying approach. The geometry, see Figure~\ref{fig:pyramids_geometry}, consists of two pyramid bodies on a build plate: the first pyramid has curved surfaces\footnote{The curvature follows a circular arc with radius $R=33 \sqrt{2}/80 \,\si{\milli\metre}$, which is tangent to the $z$-axis at the pyramid top.} and consists of bulk material, while the second pyramid is hollow. These geometries are chosen to illustrate different behavior in bulk geometries versus slender geometries. In each layer, first the bulk pyramid is scanned with a serpentine pattern with a maximal hatching space (space between two neighboring tracks' center lines) of \num{0.12} \si{\milli\metre}. Then, the hollow pyramid is scanned with a closed quadrilateral track, starting from the corner with minimal $y$- and $z$-coordinate and first moving into $y$-direction. After each processed layer, the domain is allowed to cool down for $\num{1.0}\, \si{\second}$. In the next layer, the serpentine scanning pattern in the bulk pyramid is rotated by $90$ degrees, in analogy to the tracks depicted in Figure~\ref{fig:multitrack_setup}. However, the quadrilateral track along the hollow pyramid wall always follows the same qualitative shape and only increases in size.

In terms of boundary conditions, the displacements on the bottom of the base plate are constrained to zero and the temperature is fixed to the initial temperature $T_0$. All other faces are unconstrained, i.e.,thermally insulating and not loaded by external forces. Therefore, the only cooling mechanism is heat transfer over the base plate and all energy supplied through the heat source must eventually be dissipated this way. The example geometry is still small enough that this approach is feasible and no significant convective surface heat transfer would occur in the relevant time scales of this problem.

Again, a time step of $\Delta t = \num{5e-6}\,\si{\second}$ is used for the scanning phase and the initial cooling phase of $0.002\,\si{\second}$, while the remaining $0.998\,\si{\second}$ of cooling use a time step of $\Delta t = \num{1e-3}\,\si{\second}$. Figure~\ref{fig:pyramids_geometry} gives a qualitative view of the spatial discretization. Elements are relatively undistorted and roughly equally sized. The use of a non-matching discretization between layers and between base plate and the bulk pyramid is clearly visible, and is enabled by the mortar mesh tying approach presented in Section~\ref{sec:meshtying}.

Figure~\ref{fig:double_pyramid_cuts} depicts the normal stress components as well as the von Mises equivalent stress\footnote{The von Mises equivalent stress is calculated as\\ $\sigma_{vM} = \sqrt{\sigma_{xx}^2+\sigma_{yy}^2+\sigma_{zz}^2
-\sigma_{xx}\sigma_{yy}-\sigma_{xx}\sigma_{zz}-\sigma_{yy}\sigma_{zz}
+3(\sigma_{xy}^2+\sigma_{xz}^2+\sigma_{yz}^2)}$.} after the full part was processed.
Again, the highest stresses appear close to the strongly constrained base plate. With increasing distance to the base plate the stresses relax: this happens faster in the less stiff hollow pyramid compared to the bulk pyramid. The stress distribution $\sigma_{zz}$ in vertical direction exposes a zone of compressive stresses inside the bulk pyramid and tensile stresses in the outer region, which is in agreement with the results in Section~\ref{sec:3d_single_tracks} and results in literature~\cite{Hodge2016, Li2018}. Note, that parts of the material close to the edges in the first layers were only partially molten, hence the artifacts in the stress distributions. The observations for stresses are complemented by Figure~\ref{fig:double_pyramid_displacement_warp}: here, the displacement magnitude is visualized on a deformed mesh (with deformation scaled by a factor of 5 for improved visibility). In the stiffer bulk pyramid the overall deformation is small, while the hollow pyramid can deform much easier and an undesired indentation of the side walls occurs as consequence of residual stresses.

Of course, this example could also be scaled up to more realistic part dimensions when using a layer-agglomeration approach~\cite{Hodge2016} or effective heat track- or layerwise heat sources~\cite{Zhang2018a}.

\section{Conclusion}
\label{sec:conclusion}
A thermomechanically coupled computational model for the macroscale simulation of PBFAM processes was presented. A first emphasis of the present work was the consistent modeling of material behavior in a macroscale sense. To this end, a thermo-elastic material model with an additional inelastic reference strain contribution was derived and verified with academic test cases. Crucially, this inelastic reference strain contribution, formulated in rate form, was introduced to allow for a stress-free state when solidification of molten material starts. Compared to existing (instantaneous) stress-reset procedures, which might result in jumps in the stress-field, the formulation of the reference strain term in rate form guarantees a smooth stress field. Compared to more involved material models, e.g.~\cite{Bruna-Rosso2020, Noll2020}, this simple approach can easily be integrated in standard material libraries and FEM codes while still consistently capturing the most essential aspects of residual stress generation in PBFAM. This capability has been verified in a series of elementary test cases and via reference solutions stated for the sharp-interface limit. The material model did not consider plasticity, although its inclusion is possible in a straight-forward manner as already demonstrated in the stated model equations.

For the coupling of successively processed powder layers we suggested a dual mortar mesh tying approach. Compared to other popular methods, such as element-birth or quiet-element methods, it provides additional flexibility with respect to the spatial discretization, as it allows for layerwise non-matching meshes. The applicability of both novelties to larger three-dimensional problems was demonstrated. Especially the rapid cross section changes of a rather complex geometry, involving a hollow pyramid as well as a solid pyramid with curved surfaces, illustrated the effectiveness of the proposed mortar mesh tying strategy.

\appendix

\section{Behavior of the rate-based reference strain formulation}
\label{sec:appendix_ref_strain_behavior}

For further insight into the behavior of the reference strain term, the proposed rate-based reference strain formulation \eqref{eq:epsilon_ref_rate_based} is integrated over the phase change interval to obtain
\begin{align}
    \label{eq:epsilon_ref_total_integral}
    \vepsref(r_s) = \frac{1}{r_s}\left( r_s^\text{rem}\vepsref^\text{rem} + \int_{r_s^\text{rem}}^{r_s} (\veps - \veps_T)H(\dot{r}_s)\,\dd r\right),
\end{align}
with a remaining solid fraction $r_s^\text{rem}$ at solidification start, which has a reference strain $\vepsref^\text{rem}$ from a previous solidification process, as initial value for the integration.
Here, $\heaviside{x}$ is the Heaviside step function~\eqref{eq:heaviside_definition}.

\subsection{Example scenario 1}

\begin{table}
    \centering
    \renewcommand{\arraystretch}{1.2}
    \caption{Exemplary evolution of reference strain for repeated cooling and heating assuming constant strains during solidification.}
    \label{tab:evolution_ref_strain}
    \scriptsize
    \begin{tabular}{llp{5.4cm}lll}
    \toprule
    time & temperature & reference strain & $r_p$ & $r_s$& $r_m$ \\
    \midrule
    $t^0$ & $T < T_s$ & $\vepsref^0 = 0$ &  1.0 & 0.0 &0.0\\
    $t^1$ & $T = 0.5(T_l+T_s)$ & $\vepsref^1 = 0$&  0.5 & 0.0 &0.5\\
    $t^2$ & $T < T_s$ & $\vepsref^2 = \frac{1}{0.5} 0.5 (\veps^{1} - \veps_T^{1})$& 0.5 & 0.5 &0.0\\
    $t^3$ & $T = 0.25T_l +0.75T_s$ & $\vepsref^3 = \frac{1}{0.25} 0.25 \vepsref^2 = \vepsref^2$& 0.5 & 0.25 &0.25\\
    $t^4$ & $T < T_s$ & $\vepsref^4 = \frac{1}{0.5} (0.25 \vepsref^3 + 0.25(\veps^3 - \veps_T^3)) =$\newline$ \frac{1}{2}\vepsref^3 +\frac{1}{2}(\veps^3 - \veps_T^3) $& 0.5 & 0.5 &0.0\\
    $t^5$ & $T = 0.75T_l +0.25T_s$ & $\vepsref^5 = \lim_{r_s \rightarrow 0}\frac{r_s}{r_s}\vepsref^4 = \vepsref^4$& 0.25 & 0.0 &0.75\\
    $t^6$ & $T < T_s$ & $\vepsref^6 = \frac{1}{0.75}(0\cdot\vepsref^5 + 0.75(\veps^5 - \veps_T^5)) = \veps^5 - \veps_T^5$& 0.25 & 0.75 &0.0\\
    \bottomrule
\end{tabular}
\end{table}

Table~\ref{tab:evolution_ref_strain} lists a temperature history and the corresponding reference strain evolution computed with the integral form \eqref{eq:epsilon_ref_total_integral} under the assumption that the integrand $\veps - \veps_T$ stays constant during solidification, which is a good approximation for sufficiently small phase change intervals $T_l-T_s$. At $t^0$ the material consists solely of powder phase and the reference strain is zero. After heating to $0.5(T_l+T_s)$ (50\% of material molten) at $t^1$, the material cools to a temperature below $T_s$ at $t^2$. A reference strain $\vepsref^2$ evolves during this solidification process based on the strains $\veps^1 - \veps_T^1$. Another heating up to $ 0.25T_l +0.75T_s$ (25\% of material molten) does not change the reference strain. However, note that the last term in the stress \eqref{eq:stress_total} changes continuously during melting due to the multiplication with the decreasing solid fraction (which melts before powder can melt). Not all previously created solid melts and all three phases are present at $t^3$. The next cooldown to a temperature below solidus temperature at $t^4$ leads to $\vepsref^4$, computed as a weighted average of the old reference strain $\vepsref^3$ in the remaining solid and the current strains $\veps^3 - \veps_T^3$ in the melt. As expected, the powder material does not contribute to the reference strain. A final heating to $0.75T_l +0.25T_s$ (75\% of material molten) at $t^5$ melts all of the previously created solid phase and, additionally, some more of the powder phase. Thus, the subsequent cooling at $t^6$ leads to a reference strain $\vepsref^6$, which only depends on the current strains $\veps^5 - \veps_T^5$, as no solid fraction (with a reference strain contribution) remained at $t^5$.

\subsection{Example scenario 2}
The final special case looks at cyclic, partial remelting of an initially solid material point as outlined in Table~\ref{tab:evolution_ref_strain_cyclic}. This case is also investigated as a numerical example in Section~\ref{sec:example_partial_melt}. The material is repeatedly heated up to a maximum temperature $T_m=\frac{T_s+T_l}{2}$ (i.e., 50\% of the material melts) and afterwards cooled to a temperature below $T_s$. In the first cycle, half of the initial solid will melt (at $t^1$) and obtain a new reference strain after cooldown (at $t^2$), where the strain $\veps -\veps_T$ is assumed (approximately) constant during the complete solidification process. The reference strain of the non-molten solid remains at its initial value, i.e., zero. The total reference strain after cooldown is thus $\vepsref^2=0.5({\veps} - \veps_T)$. In the next cycle, again 50\% of the solid phase melts (at $t^3$). After cooldown below $T_s$ (at $t^4$) the new reference strain is computed according to \eqref{eq:epsilon_ref_total_integral} as a weighted average of the reference strain of the non-molten solid (at $t^3$) and the new strain contribution $\veps -\veps_T$ for the remolten fraction, $\vepsref^4=0.5\vepsref^3+ 0.5(\veps - \veps_T)=0.75(\veps - \veps_T)$. The same logic applies to the next heating at $t^5$ and cooling $t^6$ and further heating-cooling cycles. Apparently, the evolution of the pre-factor in the reference strain over these cycles follows a geometric series that converges to 
${\veps} - \veps_T$. If this process is repeated many times, this has the (at first glance) paradoxical result that, although never fully molten, the complete initial solid will asymptotically convert to newly solidified solid.
From a microscale perspective this observation can be interpreted as follows: within a representative volume, the choice of the solid material fractions  (i.e., ensembles of molecules) that will actually melt is random and will change during the repeated partial melting cycles. Thus, after sufficient partial melting cycles each solid material fraction has molten at least once such that the total solid phase effectively exhibits a new stress-free reference state.

\begin{table}
    \centering
    \renewcommand{\arraystretch}{1.2}
    \caption{Exemplary evolution of reference strain for a cyclic, partial remelting of solid phase assuming $\veps - \veps_T$ is constant in all phase change intervals.}
    \label{tab:evolution_ref_strain_cyclic}
    \scriptsize
    \begin{tabular}{lllll}
    \toprule
    time & temperature & reference strain & $r_s$& $r_m$ \\
    \midrule
    $t^0$ & $T < T_s$ & $\vepsref^0 = 0$ &  1.0 & 0.0\\
    $t^1$ & $T = 0.5(T_l+T_s)$ & $\vepsref^1 = 0$& 0.5 & 0.5\\
    $t^2$ & $T < T_s$ & $\vepsref^2 = \frac{1}{1} (0.5 \vepsref^1 + 0.5 (\veps - \veps_T)) = \frac{1}{2} (\veps - \veps_T)$& 1.0 & 0.0\\
    $t^3$ & $T = 0.5(T_l+T_s)$ & $\vepsref^3 = \frac{1}{0.5} 0.5 \vepsref^2 = \vepsref^2$&  0.5 & 0.5\\
    $t^4$ & $T < T_s$ & $\vepsref^4 = \frac{1}{1} (0.5 \vepsref^3 + 0.5(\veps - \veps_T)) = \frac{3}{4}(\veps - \veps_T)$& 1.0 &0.0\\
    $t^5$ & $T = 0.5(T_l+T_s)$ & $\vepsref^5 = \frac{1}{0.5} 0.5 \vepsref^4 = \vepsref^4$& 0.5 & 0.5\\
    $t^6$ & $T < T_s$ & $\vepsref^6 = \frac{1}{1} (0.5 \vepsref^5 + 0.5(\veps - \veps_T)) = \frac{7}{8}(\veps - \veps_T)$& 1.0 &0.0\\
    \bottomrule
\end{tabular}
\end{table}

\section{Analytical reference solutions for one-dimensional {problem}}
\label{sec:appendix_stress_analytical}
This section presents analytical solutions for verification of the one-dimensional scenario described in Section~\ref{sec:validation_examples}.

\subsection{Final stress after full melt and solidification}
\label{sec:appendix_stress_analytical_full}
The homogeneous Dirichlet boundary condition and homogeneous temperature load used in Sections \ref{sec:example_full_melt} and \ref{sec:example_partial_melt} imply $\varepsilon\equiv 0$. Thus, the final stress after a full melt and cooldown to initial temperature is found from \eqref{eq:stress_total} by integrating \eqref{eq:epsilon_ref_rate_based} after a change of variables from solid fraction to temperature:

\begin{align}
    \label{eq:sigma_final_homogeneous}
    \sigma_\text{final} &= -E_s\varepsilon_\text{ref}
    =E_s \int_{0}^1 \alpha_T(T-T_0)\,\dd r_s =  -E_s \alpha_T\int_{T_l}^{T_s}\frac{T-T_0}{T_l-T_s}\, \dd T \nonumber\\ &= E_s\alpha_T\left(\frac{T_s+T_l}{2} -T_0\right)
\end{align}
where the average of solidus and liquidus temperature can be interpreted as the melting temperature $T_m:=\frac{T_s+T_l}{2}$. Result~\eqref{eq:sigma_final_homogeneous} is equivalent to the expected solution for isothermal phase change at melting temperature $T_m$ and the exact value of the (artificial) powder and melt stiffness is irrelevant for the final stress value in this specific case.

\subsection{Stress after partial melt and solidification}
\label{sec:appendix_stress_analytical_partial}

The computation in \eqref{eq:sigma_final_homogeneous} can be generalized to find the final stress after a partial melt with a peak temperature $\hat{T}$, lying in the phase change interval $[T_s;T_l]$, and subsequent cooldown:
\begin{align}
    \label{eq:sigma_final_partial_melt}
    \sigma_\text{final} &=  -E_s\varepsilon_\text{ref} =
     -E_s \alpha_T\int_{\hat{T}}^{T_s}\frac{T-T_0}{T_l-T_s}\, \dd T = \underbrace{\frac{\hat{T}-T_s}{T_l-T_s}}_{ = g(\hat{T})} \left(\frac{\hat{T}+T_s}{2}-T_0\right).
\end{align}

\subsection{Final stress for moving temperature peak}
\label{appendix:analytical_1d_inhomogeneous}

The following derivations consider the example in Section~\ref{sec:inhomogeneous_temperature}, where the space and time-dependent, peak-shaped temperature profile $T(x,t)$ given in~\eqref{eq:1d_temperature_peak_profile} is prescribed on the whole domain. We assume  zero stiffness for powder and melt phase and a finite value for the solid phase. Let $\hat{t}$ be the instant of time, when, for the first time, there is a solid fraction $r_s>0$ everywhere in the bar. No stress occurs before and at this time (due to zero stiffness in powder and melt) and consequently $\sigma(\hat{t}) = 0$.
With these simplifications we find from \eqref{eq:stress_total} 
\begin{align}
    \sigma(t) = r_s(x,t)E_s\left(\varepsilon(x,t) - \alpha_T(T(x,t)-T_0) - \varepsilon_\text{ref}(x,t)\right), \quad \forall t \geq \hat{t}
\end{align}
where the spatial dependency on $x$ is left out for the stress. Due to geometric compatibility the displacement on both ends of the domain must vanish, and therefore
 \begin{align}
    \label{eq:analytical_sol_geom_compat}
     0 &= \int_0^l \varepsilon(x,t)\, \dd x \nonumber \\ &= \sigma(t)\int_0^l\frac{\dd x}{r_s(x,t)E_s} + \int_0^l\alpha_T\left(T(x,t)-T_0\right)\, \dd x + \int_0^l \varepsilon_\text{ref}(x,t)\, \dd x.
 \end{align}
 Since $\sigma(\hat{t}) = 0$ as stated above, evaluating \eqref{eq:analytical_sol_geom_compat} at $\hat{t}$ yields
 \begin{align}
    \label{eq:analytical_sol_geom_compat_at_solidification}
     0 = \int_0^l\alpha_T\left(T(x,\hat{t})-T_0\right) \, \dd x+ \int_0^l\varepsilon_\text{ref}(x,\hat{t})\, \dd x,
 \end{align}
 and subtracting \eqref{eq:analytical_sol_geom_compat_at_solidification} from \eqref{eq:analytical_sol_geom_compat} gives
\begin{align}
    \label{eq:analytical_sol_subtracted}
    0 &= \sigma(t)\int_0^l\frac{\dd x}{r_s(x,t)E_s} + \int_0^l\alpha_T\left(T(x,t)-T(x,\hat{t})\right)\, \dd x \nonumber \\ &+ \int_0^l \varepsilon_\text{ref}(x,t) - \varepsilon_\text{ref}(x,\hat{t})\, \dd x.
\end{align}
Let $t_f$ be the final time, where all material is solid, $r_s(x, t_f)=1$, and the temperature is equal to the initial temperature, $T(x,t_f) = T_0$. Then the first term in \eqref{eq:analytical_sol_subtracted} is trivial to integrate and, after some rearrangement, one finds for the final stress:
\begin{align}
    \label{eq:1d_slab_sigma_general}
    \sigma_\text{final} = \sigma\left(t_f\right) = \frac{E_s}{l}\underbrace{\int_0^l \alpha_T \left(T(x,\hat{t})-T_0\right)\,\dd x}_{I_1} + \frac{E_s}{l}\underbrace{\int_0^l  \varepsilon_\text{ref}(x,\hat{t}) - \varepsilon_\text{ref}(x,t_f)\, \dd x}_{I_2}.
\end{align}
The first term depends on the integrated difference in temperature when solidification starts (at time $\hat{t}$) compared to the final temperature profile (at time $t_f$). For the temperature profile~\eqref{eq:1d_temperature_peak_profile}, it can be computed in analogy to \eqref{eq:sigma_final_homogeneous} and \eqref{eq:sigma_final_partial_melt} as:
\begin{align}
    \label{eq:appendix_step_4}
    I_1 = \frac{\alpha_T}{2}w\frac{(T_l-T_0)^2}{T_\text{max}-T_0}.
\end{align}
Result \eqref{eq:appendix_step_4} demonstrates that the stress after cooldown depends on the shape of the temperature profile (here described by $w$ and $T_\text{max}$) and not just the temperature difference between liquidus temperature $T_l$ and final temperature $T_f$.

To compute the second term in \eqref{eq:1d_slab_sigma_general}, the definition of the reference strain \eqref{eq:epsilon_ref_rate_based} is inserted:
\begin{align}
    \label{eq:appendix_step_5}
    I_2 &= \int_0^l  \underbrace{\varepsilon_\text{ref}(x,\hat{t})}_{=0, \text{ since for }t\leq \hat{t}:\ \varepsilon=\alpha_T\Delta T} - \varepsilon_\text{ref}(x,t_f)\, \dd x\nonumber\\ &= - \int_0^l \frac{1}{r_s(x,t_f)} \int_{\hat{t}}^{t_f} H(\dot{r}_s) \left(\varepsilon(x,\tilde{t})-\alpha_T\Delta T(x,\tilde{t})\right) \dot{r}_s(\tilde{t})\,\dd \tilde{t}\, \dd x
\end{align}
Assume that the strain stays constant during solidification, i.e.,  $\varepsilon(x,\tilde{t})=\varepsilon(x,\hat{t})=\alpha_T\Delta T(x,\hat{t})$, which is a good approximation for a small phase change interval $[T_s;T_l]$. The final solid fraction is again given as $r_s(x, t_f)=1$ and its rate during solidification as $\dot{r}_s = \dot{T}/(T_s-T_l)$. Thus, \eqref{eq:appendix_step_5} simplifies to:
\begin{align}
    I_2 = -\int_0^l \int_{\hat{t}}^{t_f} H(\dot{r}_s) \alpha_T(T(x,\hat{t}) - T(x,\tilde{t})) \frac{\dot{T}}{T_s-T_l}\,\dd\tilde{t}\, \dd x
\end{align}
After a change of variables, $H(\dot{r}_s)\dot{T}\,\dd\tilde{t} \rightarrow H\left({T}(x,\hat{t})-T_s\right) \dd \tilde{T}$, the inner integral can be computed as:
\begin{align}
  \label{eq:appendix_step_7}
    I_2 &= -\alpha_T \int_0^l H\left({T}(x,\hat{t})-T_s\right) \int_{{T}(x,\hat{t})}^{T_s}  \frac{\tilde{T} - {T}(x, \hat{t})}{T_l-T_s}\,\dd \tilde{T}\,\dd x \nonumber\\ &= -\frac{\alpha_T}{2} \int_0^l H\left({T}(x,\hat{t})-T_s\right) \frac{(T_s-{T}(x,\hat{t}))^2}{T_l-T_s}\,\dd x
\end{align}
Finally, the specific temperature profile \eqref{eq:1d_temperature_peak_profile} is inserted into  \eqref{eq:appendix_step_7} which allows to compute the final integral, yielding
\begin{align}
    I_2 = -\frac{\alpha_T}{6} w \frac{(T_l-T_s)^2}{T_\text{max}-T_0},
\end{align}
which is a significantly smaller value compared to $I_1$ for typical values encountered in PBFAM application, e.g. for the parameters used in Section~\ref{sec:inhomogeneous_temperature} a ratio  of $\vert I_2/I_1 \vert \approx 0.3\%$ is obtained.

\section*{Acknowledgements}
This work was supported by the German Research Foundation (DFG) and the Technical University of Munich (TUM) in the
framework of the Open Access Publishing Program.

\section*{Funding}
This work was supported by funding of the Deutsche Forschungsgemeinschaft (DFG, German Research Foundation) within project
437616465 and project 414180263.

\bibliographystyle{abbrv}
\bibliography{ref}
\end{document}